\documentclass[pre,aps,twocolumn,showpacs,floatfix,superscriptaddress]{revtex4-2}

\usepackage{graphicx}
\usepackage{bm}
\usepackage{color}
\usepackage{amssymb,amsmath}
\usepackage{ulem}
\usepackage{ragged2e}
\usepackage{adjustbox}
\usepackage{multirow}
\usepackage{booktabs}
\usepackage[table]{xcolor}
\usepackage{tcolorbox}
\usepackage{tabularx}
\usepackage{array}
\usepackage{colortbl}
\usepackage{hyperref}
\hypersetup{colorlinks=true, citecolor=blue, urlcolor=blue, linkcolor=blue}
\tcbuselibrary{skins}
\definecolor{Salmon}{RGB}{250,128,114}

\newcolumntype{Y}{>{\raggedleft\arraybackslash}X}

\tcbset{tab1/.style={\centering, fonttitle=\bfseries\small,fontupper=\normalsize\sffamily,
		colback=yellow!10!white,colframe=red!75!black,colbacktitle=Salmon!40!white,
		coltitle=black,center title,freelance,frame code={
			\foreach \n in {north east,north west,south east,south west}
			{\path [fill=red!75!black] (interior.\n) circle (3mm); };},}}

\tcbset{tab2/.style={enhanced,fonttitle=\bfseries,fontupper=\small\sffamily,
		colback=yellow!10!white,colframe=red!50!black,colbacktitle=Salmon!40!white,
		coltitle=black,center title}}

\begin{document}

\title{Extreme events in two-coupled chaotic oscillators}
\author{S. Sudharsan}
\affiliation{Physics and Applied Mathematics Unit, Indian Statistical Institute, Kolkata 700108, India}
\author{Tapas Kumar Pal}
\affiliation{Physics and Applied Mathematics Unit, Indian Statistical Institute, Kolkata 700108, India}
\affiliation{Department of Mathematics, Jadavpur University, Kolkata 700032, India}
\author{Dibakar Ghosh}
\affiliation{Physics and Applied Mathematics Unit, Indian Statistical Institute, Kolkata 700108, India}
\author{J\"{u}rgen Kurths}
\affiliation{Potsdam Institute for Climate Impact Research - Telegraphenberg A 31, Potsdam, 14473, Germany}
\affiliation{Humboldt University Berlin, Department of Physics, Berlin, 12489, Germany}

\begin{abstract}
Since 1970, the R\"ossler system has remained as a considerably simpler and minimal dimensional chaos serving system. Unveiling the dynamics of a system of two coupled chaotic oscillators that leads to the emergence of \textit{extreme events} in the system is an engrossing and crucial scientific research area. Our present study focuses on the emergence of \textit{extreme events} in a system of \textit{diffusively} and \textit{bidirectionally} two coupled R\"ossler oscillators and unraveling the mechanism behind the genesis of \textit{extreme events}. We find the appearance of \textit{extreme events} in three different observables: average velocity, synchronization error, and one transverse directional variable to the synchronization manifold. The emergence of \textit{extreme events} in average velocity variables happens due to the occasional \textit{in-phase} synchronization. The \textit{on-off intermittency} plays for the crucial role in the genesis of \textit{extreme events} in the synchronization error dynamics and in the transverse directional variable to the synchronization manifold. The bubble transition of the chaotic attractor due to the \textit{on-off intermittency} is illustrated for the transverse directional variable. We use generalized extreme value theory to study the statistics of extremes. The \textit{extreme events} data sets concerning the average velocity variable follow generalized extreme value distribution. The inter-event intervals of the \textit{extreme events} in the average velocity variable spread well exponentially. The upshot of the interplay between the coupling strength and the frequency mismatch between the system oscillators in the genesis of \textit{extreme events} in the coupled system is depicted numerically.

\end{abstract}
\maketitle
\section{Introduction}

The rudiments of nature make every living creature beings a far potential of having survivability. Several ubiquitous, catastrophic natural and human-made challenges drive human spontaneous endeavors to consecrate themselves for having an outlook of being capable of surviving. Natural calamities like epidemic spreading \cite{li2021emergence}, floods \cite{sundaram2021modeling}, earthquakes \cite{qi2024seismic}, droughts \cite{giesche2023recurring}, regime shifts in the ecosystem \cite{andisa2023ecosystem}, global warming \cite{osman2023planting}, cyclones \cite{boragapu2023tropical}, and tsunamis \cite{ROBKE2017296} to name but a few, and several human-made disasters like power blackouts \cite{suncascading}, nuclear leakage \cite{bona2023nuclear}, crash in the share market \cite{Sornette+2003} have a terrible ruinous impact on the human socio-economic structure \cite{sayantan2021review}. Beholding the uncertainty to predict the future spontaneous occurrence of low probable recurrent natural and human made hazards having immediate severe, harsh consequences on the society, a new trend of fascinating research interest of \textit{extreme events} (EEs) has been started in many interdisciplinary disciplines of scientific research \cite{sayantan2021review,farazmand2019review,mcphillips2018review,WALSH2020103324,ummenhofer2017review,stewart2022review}. 
Generally, the events or phenomena having a significant deviation from the regular behavior with a colossal impact in terms of havoc on the society are regarded as EEs \cite{farazmand2019review,sayantan2021review}. These types of phenomena have synergistic consequences of the abrupt changes in the system states.

\par EEs are characterized as the occurrence of events far distant from the central tendency of the events \cite{farazmand2019review}. Depending on the deviation how far it is from the central tendency \cite{dysthe2008hs}, rarity of EEs is recognized \cite{kantz2016general}. 
EEs being low probable to occur, their appearance is seen on the tail of the Non-Gaussian skewed distribution \cite{lucarini2016extremes,ghill2011review}. Prediction \cite{SILLMANN201765} and obviation \cite{osman2023planting} is the main concern to study EEs. As far as the real world scenario \cite{mcphillips2018review,stewart2022review} is concerned, it is an utmost challenge to the scientific community to examine EEs and infer their grievous impact on society due to the paucity of observed data \cite{altwegg2017single}. In this regard, the researchers whet their appetite towards the dynamical systems \cite{sayantan2021review}. Numerically simulating the dynamical system, one can gather enormous data which palliate the scarcity of the real data. The main reason of origination of EEs in dynamical systems is the presence of \textit{instability region} within the state space. Seldom visit of a chaotic trajectory in the \textit{instability} region aftermaths a long excursion of the trajectory far from the chaotic attractor for a short while and follows a return back \cite{farazmand2019review}. This long excursion is observed as unwonted events of large amplitude in the time series of the system. The emergence of EEs in dynamical systems mainly happened following three instability regions originating roots: \textit{interior-crisis-induced} \cite{kingston2017extreme,mishra2020route,Sudharsan_2021,ray2019extreme}, \textit{intermittency-induced} \cite{kingston2017extreme,mishra2020route,Sudharsan_2021}, and \textit{quasiperiodic-breakdown} \cite{mishra2020route}.  
In multistable systems, the trajectory may start hopping between the coexisting stable states as a consequence, episodic large-amplitude events may emerge in the dynamical system. Other routes, like boundary crisis and attractor merging crisis are also observed  in other systems.
\par The genesis of \textit{extreme events} in single (uncoupled) nonlinear dynamical systems and the possible mechanisms \cite{sayantan2021review,ray2019intermittent,ray2020understanding,pal2023extreme, sudharsan2021emergence,Sudharsan_2021,kingston2017extreme,kumarasamy2018extreme,kaviya2022extreme,kaviya2023route,thangavel2021extreme,manivelan2024dynamical} behind the origination of EEs have almost been unfolded. Comparatively, the responsible mechanisms behind the origination of EEs in \textit{coupled} dynamical systems are less studied. So, the unveiling of mechanisms behind the emergence of high-amplitude, erratic EEs in coupled, both in low-dimensional \cite{ansmann2013extreme,saha2017extreme,mishra2018dragon,kumarasamy2021travelling,vijay2024transition} and high-dimensional \cite{ansmann2013extreme,karnatak2014extreme,kumarasamy2021travelling,kumarasamy2022time,kingston2023timevary,ray2020josephson,ray2022networks,kanagaraj2024unraveling,roy2023extreme,chowdhury2019synchronization,chowdhury2021extreme}, dynamical systems have been a hot spot of interdisciplinary scientific research for the last few years. In this direction, some notable scientific research has been performed, like the emergence of EEs in networks of Fitzhugh-Nagumo oscillators \cite{ansmann2013extreme}, the genesis of EEs in a globally coupled network of Josephson junction oscillators and Li\'enard type oscillators \cite{ray2020josephson}.
Ray \textit{et al.} \cite{ray2022networks} have shown that the interplay of degree heterogeneity and repulsive interaction produces extreme events in a complex network of second-order phase oscillators. Time-dependent interactions and rewiring of the links of a network have also been found as a precursor for the origination of extreme events \cite{kumarasamy2022time,kingston2023timevary,roy2024extreme}.
 Moreover, network changes such as introduction of time-delayed coupling \cite{saha2017extreme}, pairing of two-counter rotating oscillators \cite{kumarasamy2021travelling} have also been shown to produce extreme events.
 \par Till now, the unearthing of mechanisms behind the nascence of EEs in coupled dynamical systems is in its infancy. In this present study, a \textit{diffusively} and \textit{bidirectionally} two-coupled R\"ossler oscillator system is considered. This model has already been studied concerning chaotic phase synchronization (CPS) \cite{osipov2003three}, but this condign study focusses on comprehending the exploration of the emergence of extreme events in the same system. Our main concern in the study is to investigate the system's dynamical behavior in the presence of frequency mismatch and discern whether it leads to the genesis of EEs in the system and the crucial dynamical mechanism behind it regarding the interplay between the frequency mismatch and the coupling strength. Interestingly, we observed the emergence of EEs in the system for three observables, {$u=\frac{\dot{x}_1+\dot{x}_2}{2}$} (the average velocity variable in the $x$ direction), the synchronization error ($E_{syn} =\langle \sqrt{(\dot{x}_1-\dot{x}_2)^2+(\dot{y}_1-\dot{y}_2)^2+(\dot{z}_1-\dot{z}_2)^2}\rangle_t$), and one transverse directional variable ($(x_{\perp})_{3}=\frac{\dot{z_{1}}-\dot{z_{2}}}{2}$) to the synchronization manifold. We unraveled in detail the dynamical mechanisms involved in the origination of EEs. The emergence of EEs happens in the observable {$u=\frac{\dot{x}_1+\dot{x}_2}{2}$} because of \textit{occasional in-phase synchronization} of the variables $\dot{x}_1$ and $\dot{x}_2$. The genesis of EEs in the synchronization error dynamics and in the transverse directional variable $(x_{\perp})_{3}=\frac{\dot{z_{1}}-\dot{z_{2}}}{2}$ happens following \textit{on-off intermittency} route. So far as our knowledge is concerned, this is the first time that we are reporting this kind of emergence of EEs in a system of two coupled R\"ossler oscillators. For corroboration of the results, some statistical analyses are also performed. An illustration is done for two specific coupling strengths that the EEs in $u$ follow a non-Gaussian generalized extreme value (GEV) distribution, confirming the rarity. The inter-event intervals (IEI) are also shown to follow a non-Gaussian exponential distribution that corroborates the rare occurrence. The elucidation of how the interplay between the frequency parameter mismatch ($\Delta\omega$) and the coupling strength ($k$) influences the emergence of EEs in the system is also shown by some parameter spaces upon the $(k,\Delta\omega)$ plane.

\section{Model description}   %and Dynamics}
\label{sec:2}
%For our numerical study, we consider 
The R\"ossler oscillator, which is being well-known and prominent one, as a three-dimensional simplest system serving chaos and rendering exemplary chaotic dynamics. Our main concern of the study is to investigate how a system of two coupled R\"ossler oscillators behaves dynamically, specifically if the coupling is \textit{diffusive} and \textit{bidirectional}. So, we consider a system of two-coupled R\"ossler oscillators \cite{rosenblum1996phase,rosenblum1997phase,osipov2003three}, precisely, \textit{diffusively}, and \textit{bidirectionally} coupled in the $y$ variable. The mathematical form of the system is presented by the following system of equations.
\begin{eqnarray}
	\begin{aligned}
	\dot{x}_{1,2} &= -\omega_{1,2} y_{1,2} - z_{1,2},\\   %\nonumber \\
	\dot{y}_{1,2} &= \omega_{1,2} x_{1,2} + ay_{1,2} + k(y_{2,1}-y_{1,2}),\\   %\nonumber \\
	\dot{z}_{1,2} &= b + z_{1,2}(x_{1,2} - c),
	\end{aligned}
	\label{mastereqn}
\end{eqnarray}
where $k$ is the coupling strength of the interaction between the two oscillators. %Throughout the study, the 
Numerical simulations are performed using the Runge-Kutta Fehlberg (RKF) 45 algorithm with a fixed step size of $0.01$ considering $12\times10^7$ iterations and abjuring $9\times10^7$ transients, and all the parameter values remain fixed at $a=0.24$, $b=0.1$, $c=8.5$, and $\omega_{1,2}=1.0\pm \Delta\omega$, where $\Delta\omega$ creates a slight mismatch of the frequencies between the two oscillators. 
%and its numerical value is presented at the appropriate place within the manuscript. 
For the numerical integration, $(x_{1,2}(0),y_{1,2}(0),z_{1,2}(0)) \approx (0.10,0.05,0.15)$ is considered as the initial condition in every case. For the system \eqref{mastereqn}, we define three new variables \textbf{$u = \dot{x}_{avg}=\frac{\dot{x}_1+\dot{x}_2}{2}$}, \textbf{$v = \dot{y}_{avg}=\frac{\dot{y}_1+\dot{y}_2}{2}$} and \textbf{$w = \dot{z}_{avg}=\frac{\dot{z}_1+\dot{z}_2}{2}$} for scrupulous observation of the collective dynamics of the comprising oscillators.
\textit{Measure of extremes}: Hitherto, there is no concordant scientific definition of \textit{extreme events} or \textit{extremes} in the literature. The events that deviate significantly from the central tendency for a specific set of events are conventionally contemplated as EEs. The very term significantly deviated is pivotal. There are several statistical techniques available in the literature to define the \textit{significant deviation}, like the 90th–99th percentile of the probability distribution of the respective set of events and the significant height, or threshold, based on the set of events.

\par As far as the study of EEs in dynamical systems is concerned, the significant height or threshold-based method to classify EEs is a rife one. If $\mu$ is the mean and $\sigma$ is the standard deviation of a set of events, then the threshold for this very set is defined as $H_{th}=\mu + d\sigma,\,d\in\mathbb{R}\setminus\{0\}$. The events that surpass this threshold $H_{th}$ are appraised as EEs. The numerical value of $d$ discerns how far an event is deviated from the mean state of the data set of events. The large numerical value of $d$ makes sense for the rarity of EEs. Throughout the study, we consider the \textit{threshold-based} approach to classifying the EEs, and we choose $d=-5$ to define the threshold, i.e., in our case the threshold is $H_{th}=\mu - 5\sigma$.
%$\dot{x}$ is the observable and
\par In this present study, \textbf{$u = \dot{x}_{avg}=\frac{\dot{x}_1+\dot{x}_2}{2}$} is the \textit{observable}, and the local minima of $u$, i.e., $u_{min}$ is the concerning \textit{events}, and the events that exceed the threshold $H_{th}=\mu - 5\sigma$ in the negative direction (below) are reckoned as EEs.

\section{Result}
\label{sec:4}
\begin{figure}[h!]
	\centering
	\includegraphics[width=1.0\linewidth]{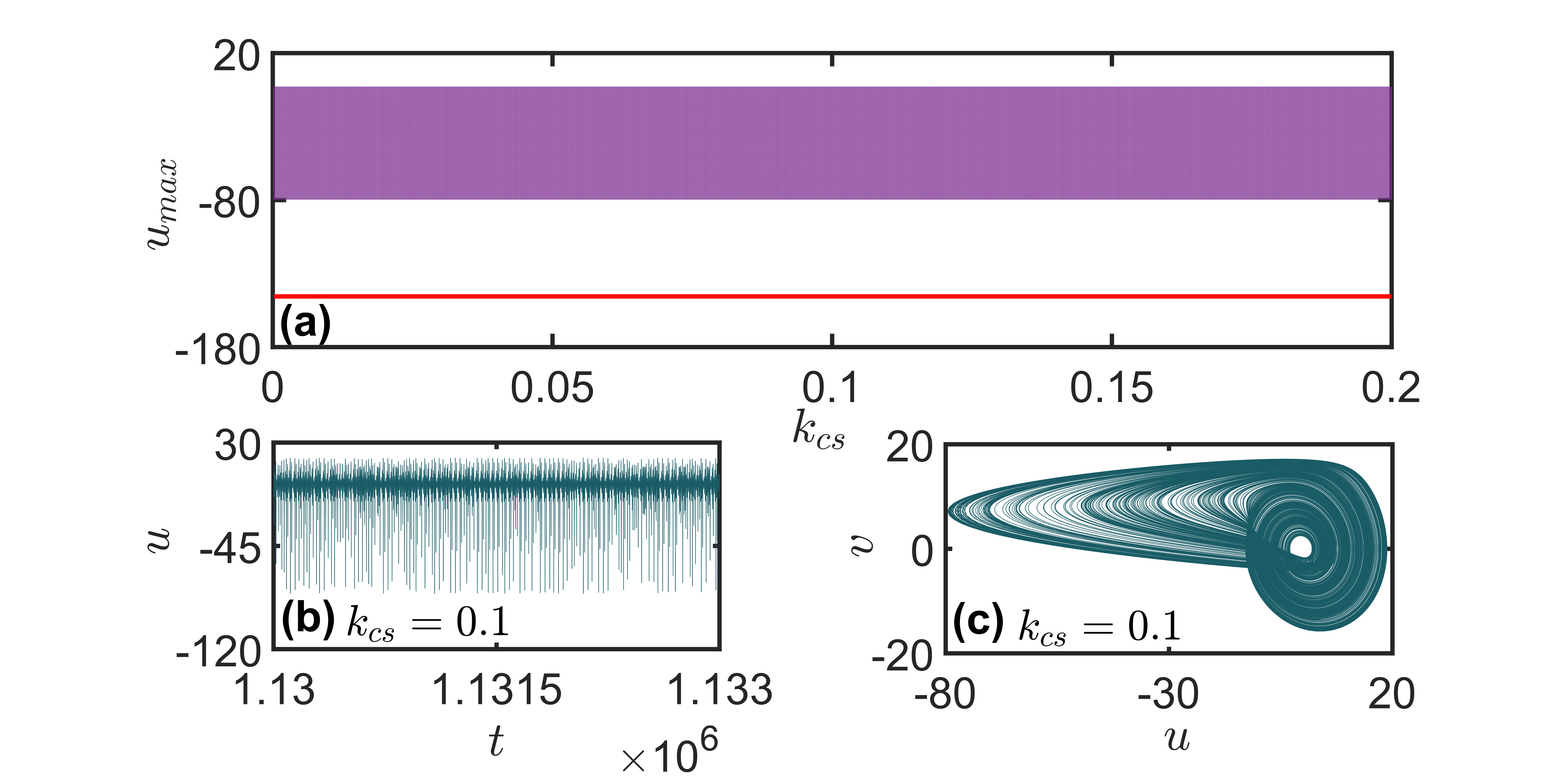}
	%\captionsetup{justification=justified} % Caption justified in two columns
	\caption{\textbf{Dynamics of the system (\ref{mastereqn}) for $\Delta\omega=0$}: (a) The changing scenario of $u_{min}$ for the variation of the coupling strength $k$ in the range $[0,0.2]$. (b) Time series for $k=0.1$. (c) Phase portrait for $k=0.1$. The result shows no EEs for identically coupled oscillators.} 	
	\label{bifur_i}
\end{figure}

%\subsection{Dynamics for identical frequency $\Delta\omega=0$}
\textit{Dynamics in the absence of frequency mismatch} ($\Delta\omega=0$): We investigate how the system \eqref{mastereqn} behaves dynamically, in particular, when $\Delta\omega=0$, i.e., for identical frequency. Considering $\omega_{1,2}=1.0$, the changing scenario of the events, i.e., $u_{min}$ with respect to the coupling strength $k$ varying in the range $[0,0.2]$, is portrayed in Fig.~\ref{bifur_i}(a), and the red line represents the threshold $H_{th}$. Figure~\ref{bifur_i}(a) exquisitely elucidates bounded chaos throughout the range $[0,0.2]$ of the coupling strength $k$, and no portion of the chaotic region surpasses the threshold line, i.e., no presence of EEs \textit{region}. For a specific value of the coupling strength $k\approx0.1$, time evaluation of the variable $u$, and phase portrait of the system \eqref{mastereqn} upon the $(u,v)$ plane are demonstrated in Fig.~\ref{bifur_i}(b) and Fig.~\ref{bifur_i}(c), respectively.

\begin{figure*}[!ht]
	\centering
	\includegraphics[width=1.01\linewidth]{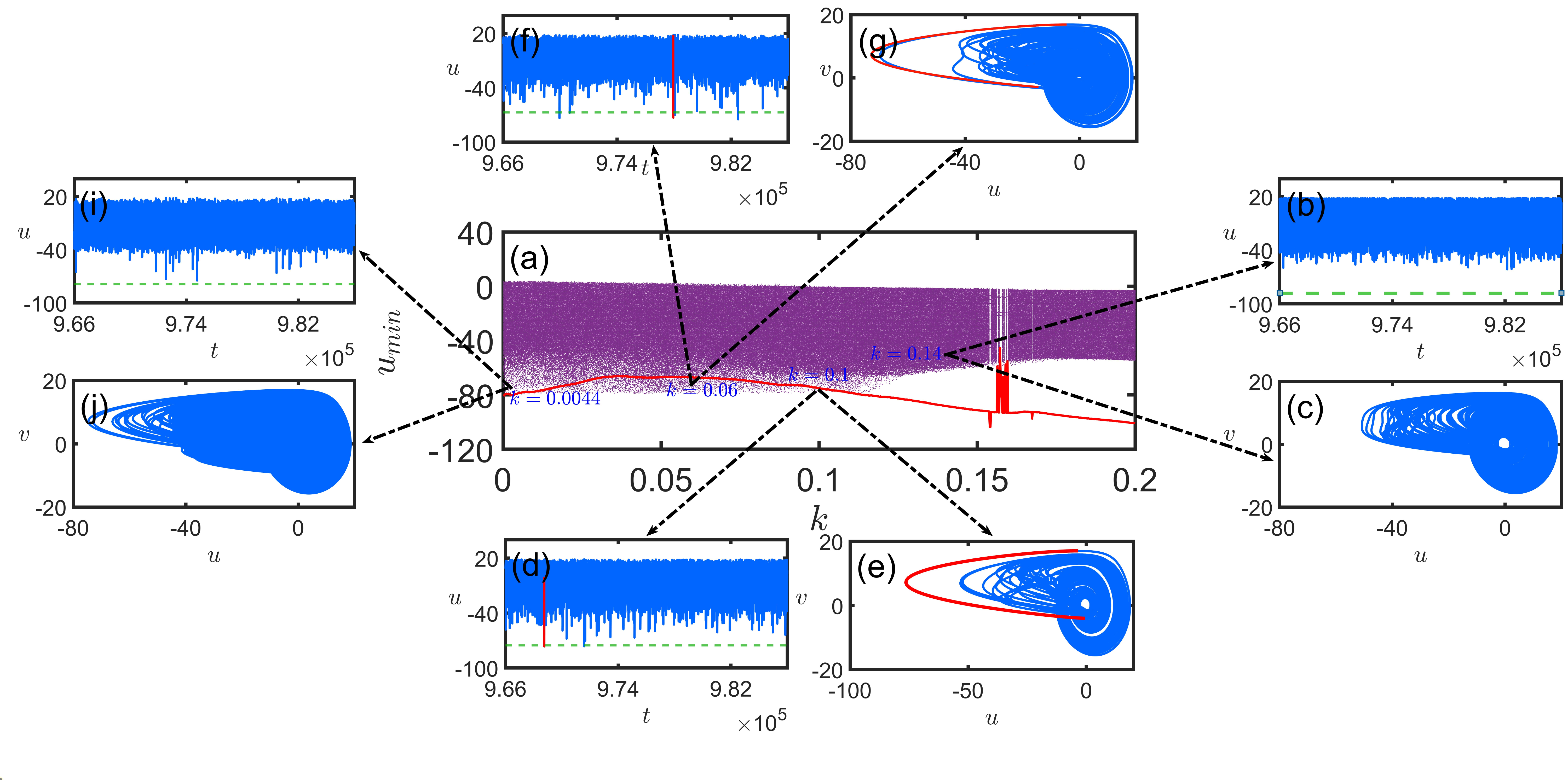}
	%\captionsetup{justification=justified} % Caption justified in two columns
	\caption{\textbf{Dynamical scenario of the system \eqref{mastereqn}:} (a) The \textit{bifurcation} diagram of $u_{min}$ is portrayed, taking the coupling strength $k$ as the \textit{bifurcation parameter}, and varying in the range $[0,0.2]$. The red curve represents the threshold, $H_{th}=\mu - 5\sigma$. Below the threshold, $H_{th}$, line, the portion of the \textit{bifurcation} diagram is the region of \textit{extreme events}. (b) The time series of the observable $u$ for $k\approx0.14$, the \textit{green-dashed} line represents the threshold, $H_{th}$, line. It is noticeable that no spike is crossing the threshold line. (c) The phase portrait for $k\approx0.14$ upon the $(u,v)$ plane. (d) The temporal evolution of $u$ for $k\approx0.1$, it is clearly recognizable that few spikes cross the \textit{green-dashed} threshold line, which are being enunciated as EEs; one such spike is shown by the \textit{red} line, and the corresponding trajectory is shown by the \textit{red} curve in the respective phase portrait upon the $(u,v)$ plane, displayed in (e). (f) The temporal evolution for $k\approx0.06$, some spikes exceed the EEs qualifying threshold line $H_{th}$ and one such spike is highlighted by \textit{red} color. (g) The phase portrait for $k\approx0.06$ upon the $(u,v)$ plane, and the highlighted \textit{red-colored} curve represents the same spike shown by the \textit{red} color in (f). (i) Temporal evolution of $u$, and clearly no spike surpasses the EE qualifying \textit{green-dashed} threshold line. (j) Phase portrait for $k\approx0.0044$ upon the $(u,v)$ plane.
	}
	\label{bifur}
\end{figure*}

\textit{Dynamics in the presence of frequency mismatch} ($\Delta\omega\neq0$) \textit{and the advent of extreme events}: The introduction of a bit of heterogeneity in the frequency parameter $\omega$ of the system \eqref{mastereqn} drastically changes the dynamics of the very system. We introduce $\Delta\omega=0.02$, specifically $\omega_{1}=0.98$, and $\omega_{2}=1.02$. This section is mainly devoted to the investigation of how the observable $u$ behaves if the coupling strength $k$ varies in the range $[0,0.2]$. A changing scenario of $u_{min}$ as the coupling strength, $k$, varies in the range $[0,0.2]$ is delineated in Fig.~\ref{bifur}(a) as the \textit{bifurcation} diagram. 
The red line represents the threshold line $H_{th}$. As the coupling strength, $k$, decreases from the right to the left, the emergence of comparatively large amplitude bounded chaos from a little bit low amplitude bounded chaos is prominently noticeable. A region starting from $k\approx0.11$ and ending at $k\approx0.0046$ in the bifurcation diagram is discernible as being below the threshold, $H_{th}$ line. This region is basically the EEs emerging region. For the sake of clarity and brevity of the numerical investigation, we consider four specific points from four different regions of the bifurcation diagram: one from the prior EEs emerging region, $k\approx0.14$; two points from the EEs region, $k\approx0.1$, and $k\approx0.06$; one point after the EEs region, $k\approx0.0142$; and study the dynamical nature of the system for these very points. The temporal evolution of the observable $u$ for $k\approx0.14$ is portrayed in Fig.~\ref{bifur}(b). The green dashed-horizontal line represents the corresponding threshold, and no spike in the time series is observed to surpass the threshold line, as the numerical value of $k$ is chosen from the non-extreme events region. The chaotic phase portrait upon the $(u,v)$ plane for $k\approx0.14$ is displayed in Fig.~\ref{bifur}(c). Figure~\ref{bifur}(d) shows the temporal evolution of $u$ for $k\approx0.1$, and it is noticeable from the temporal dynamics that a few spikes exceed the green-dashed threshold line, which indeed represents EEs. One such extreme event is presented by \textit{red-colored} spike in Fig.~\ref{bifur}(d), and the according phase portrait upon $(u,v)$ plane is presented in Fig.~\ref{bifur}(e) and the respective \textit{extreme-trajectory} is illustrated by the \textit{red} color. The temporal evolution of $u$ for $k\approx0.06$ is drawn in Fig.~\ref{bifur}(f), here it is also recognizable that few spike are crossing the \textit{green-dashed} threshold, $H_{th}$, line, which are basically EEs. We mentioned one EE by \textit{red-colored} spike in the time series, and the corresponding \textit{extreme-trajectory} is displayed by \textit{red} color in the corresponding phase portrayed, Fig.~\ref{bifur}(g), drawn upon $(u,v)$ plane. The point $k\approx0.0044$ being chosen beyond the EEs region, no spike surpassing the \textit{green-dashed} threshold line is recognizable in the temporal evolution presented in Fig.~\ref{bifur}(i). The respective phase portrait is drawn in Fig.~\ref{bifur}(j).

\begin{figure}[h!]
	\centering
	\includegraphics[width=0.5\textwidth]{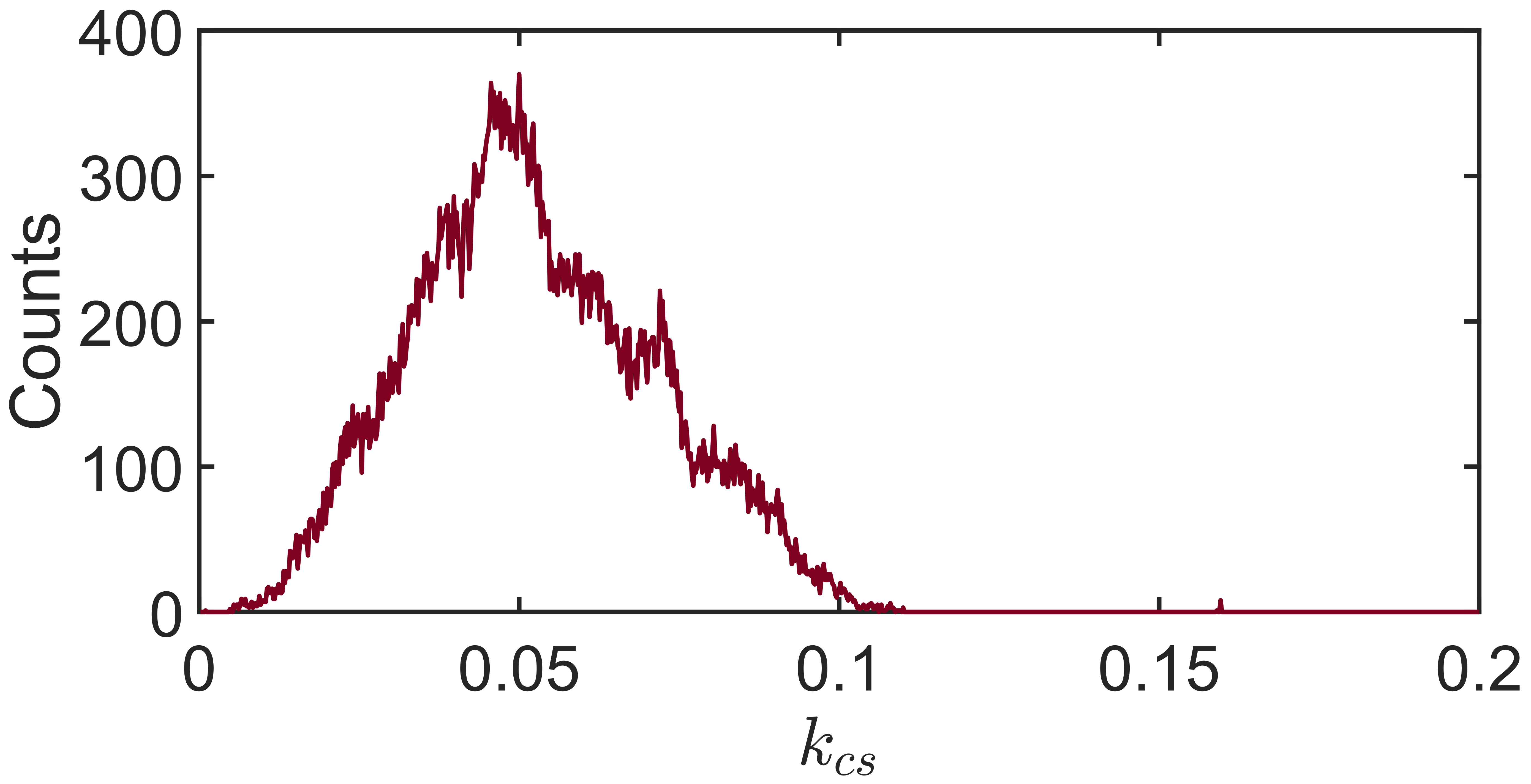}
	%\captionsetup{justification=justified} % Caption justified in two columns
	\caption{\textbf{Coupling strength vs. number of extreme events:} To substantiate the region $k\in[0.0046,0.11]$ as the extreme events merging region, a diagram is plotted here that shows, as the coupling strength $k$ varies in the range $[0,0.2]$, the number of conspicuous EEs. As $k$ decreases from the right to the left, an observation is cognizable that at $k\approx0.011$ the emergence of EEs starts and reaches its maximum number of $370$ at $k\approx0.05$, then gradually decreases, and after $k\approx0.0046$ no EE is perceivable.
	}
	\label{count}
\end{figure}

\begin{figure}[!ht]
	\centering
	\includegraphics[width=1.0\linewidth]{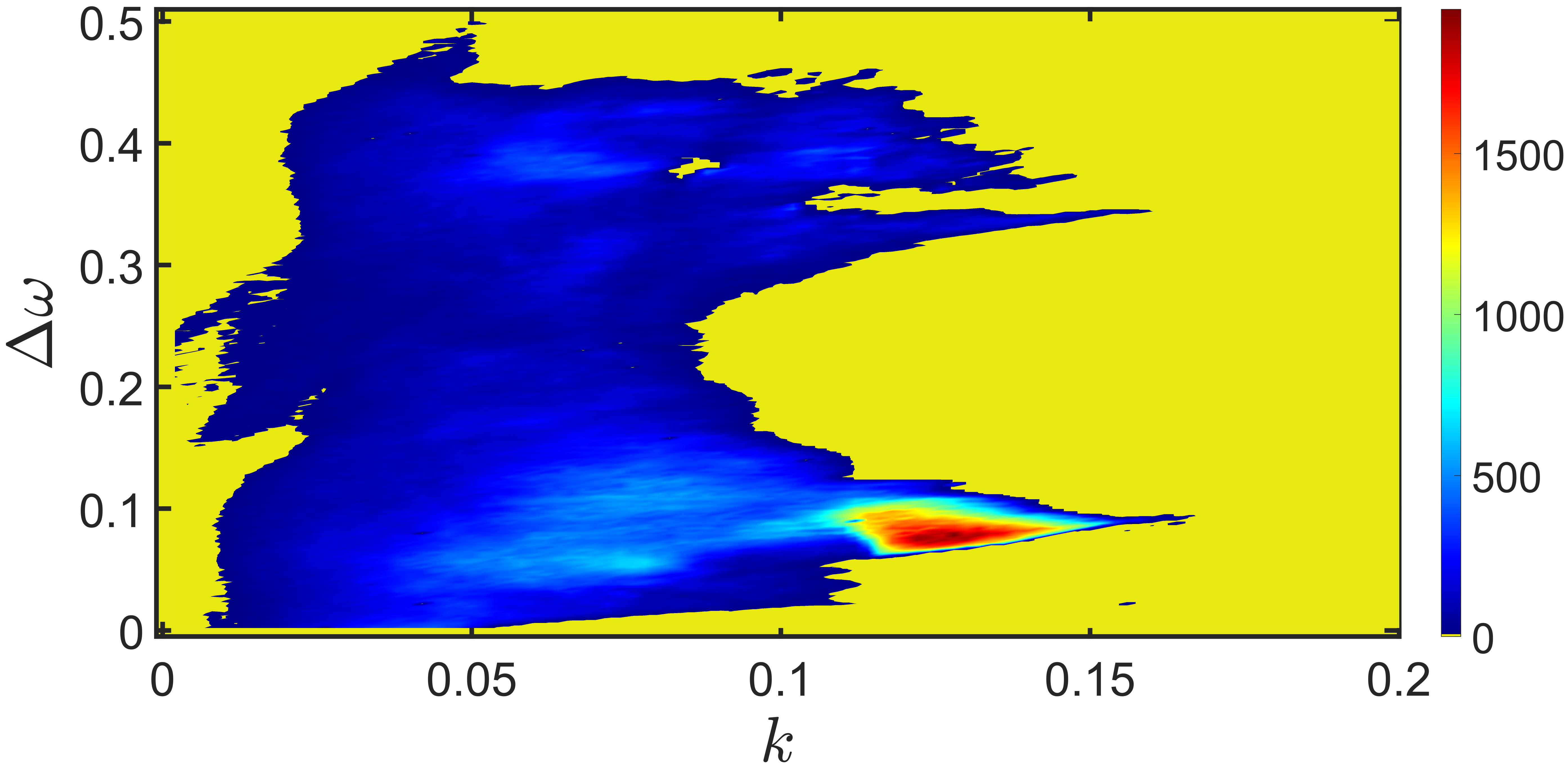}
	%\captionsetup{justification=justified} % Caption justified in two columns
	\caption{\textbf{Depiction of the frequency mismatch ($\Delta\omega$) range with the variation of the coupling strength ($k$):} For the emergence of the EEs in the system \eqref{mastereqn}, how the range of frequency mismatch ($\Delta\omega$) between the two oscillators is varied with the variation of the coupling strength ($k$) is presented as a parameter space.}
	\label{phasediag}
\end{figure}

To corroborate the region of EEs, we plot the diagram, the number of extreme events versus the coupling strength $k$, in Fig.~\ref{count}. Upon inferring the Fig.~\ref{count}, if we proceed from the right end towards the left, it is perceptible 
that as the value of $k$ decreases, the number of extreme events becomes non-zero at $k\approx0.11$, and a gradual increment in $k$ depicts a maximum number of EEs as 370 at $k\approx0.05$. Later on, it starts decreasing and succumbs to zero at $k\approx0.0046$.
To envisage how the interplay between the frequency mismatch ($\Delta\omega$) of the two oscillators and the coupling strength ($k$) of the system \eqref{mastereqn} entices the emergence of EEs. We plotted in Fig.~\ref{phasediag} a parameter space upon the ($k$,$\Delta\omega$) plane, where the yellow region exposes the non-EEs region, and the colored region unveils the EEs region. 
Inferring a careful observation upon Fig.~\ref{phasediag}, it whets to unravel that for lesser values of the coupling strength ($k\approx0.0046$), extreme events appear for a comparatively large range of $\Delta\omega$ varying in $[0,0.2]$, and as the value of $k$ increases, the range of $\Delta\omega$ gradually decreases for the occurrence of EEs.
\section{Statistical analysis}
\label{sec:5}

\begin{figure}[h!]
	\centering
	\includegraphics[width=0.5\textwidth]{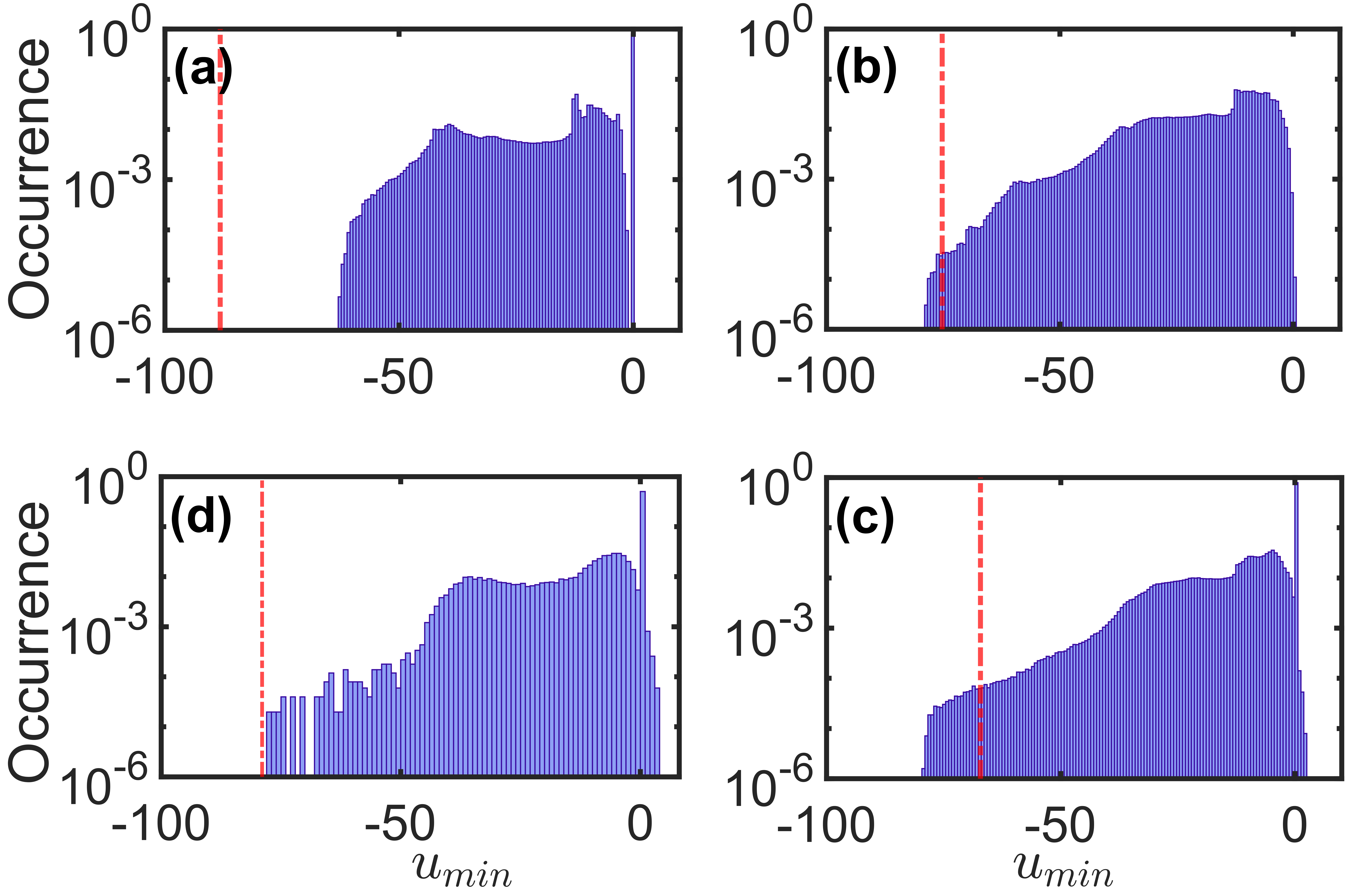}
	%\captionsetup{justification=justified} % Caption justified in two columns
	\caption{\textbf{Histogram plot:} The histograms of the numerical values of the \textit{events} ($u_{min}$) for four different values of the coupling strength ($k$) are presented here in semi-log scale. The \textit{red-dashed} vertical lines in all the figures represent the threshold, $H_{th}=\mu-5\sigma$, where $\mu$ is the mean and $\sigma$ is the standard deviation of the respective \textit{events} sets. (a) The histogram of the \textit{events} for $k\approx0.14$. No portion of the histogram crossed the \textit{red} vertical dashed threshold line, towards the left that ratifies the presence of no EE. (b) Histogram of the \textit{events} for $k\approx0.1$. The portion of the histogram on the left of the \textit{red-dashed} vertical $H_{th}$ line depicts the EEs. (c) Presentation of the histogram of the \textit{events} for $k\approx0.06$. The EEs are being accredited by the left portion of the vertical \textit{red-dashed} $H_{th}$ line. (d) Histogram of \textit{events} for $k\approx0.0044$. The non-presence of the histogram on the left of the vertical \textit{red-dashed} $H_{th}$ line apprehends the scenario of having no EE.  
	}
	
	\label{occurrences}
\end{figure}

To endorse the characteristics presented in the time series in Fig.~\ref{bifur}(b),(d),(f), and(i) statistically, we delineated the histogram plots of \textit{events} ($u_{min}$) corresponding to each time series in Fig.~\ref{occurrences}. The \textit{red} vertical dashed line plays for the threshold $H_{th}=\mu-5\sigma$. For $k\approx0.14$, the histogram of the \textit{events} corresponding to the time series presented in Fig.~\ref{bifur}(b) is displayed in Fig.~\ref{occurrences}(a). The non-exceedance of the histogram beyond the vertical $H_{th}$ line towards the left corroborates the non-presence of EEs in the respective time series. The histogram of the \textit{events} for $k\approx0.1$ is delineated in Fig.~\ref{occurrences}(b), and the left side portion of the vertical \textit{red-dashed} threshold line is consonant with the EEs spikes, which indeed surpassed the \textit{green-dashed} $H_{th}$ line. Figure~\ref{occurrences}(c) is the depiction of the histogram of the \textit{events} concerning the temporal evolution presented in Fig.~\ref{bifur}(f). The left side portion of the \textit{red} vertical dashed $H_{th}$ line of the histogram reinforce the appearance the EEs. Figure~\ref{occurrences}(d) is the rendition of the histogram of the \textit{events} regarding the time evolution displayed in Fig.~\ref{bifur}(i) corresponding to the coupling strength $k\approx0.0044$. As no EE is depicted in the time series, no portion of the histogram is discernible on the left of the $H_{th}$ line. For the statistical analysis, all the numerical simulations are carried out using the RKF45 algorithm, considering $10^9$ iterations and initially renouncing $10^5$ iterations as transient, and for performing the numerical integration, $0.01$ is considered as step length.  

\begin{figure*}[!ht]
	\centering
	\includegraphics[width=1.0\textwidth]{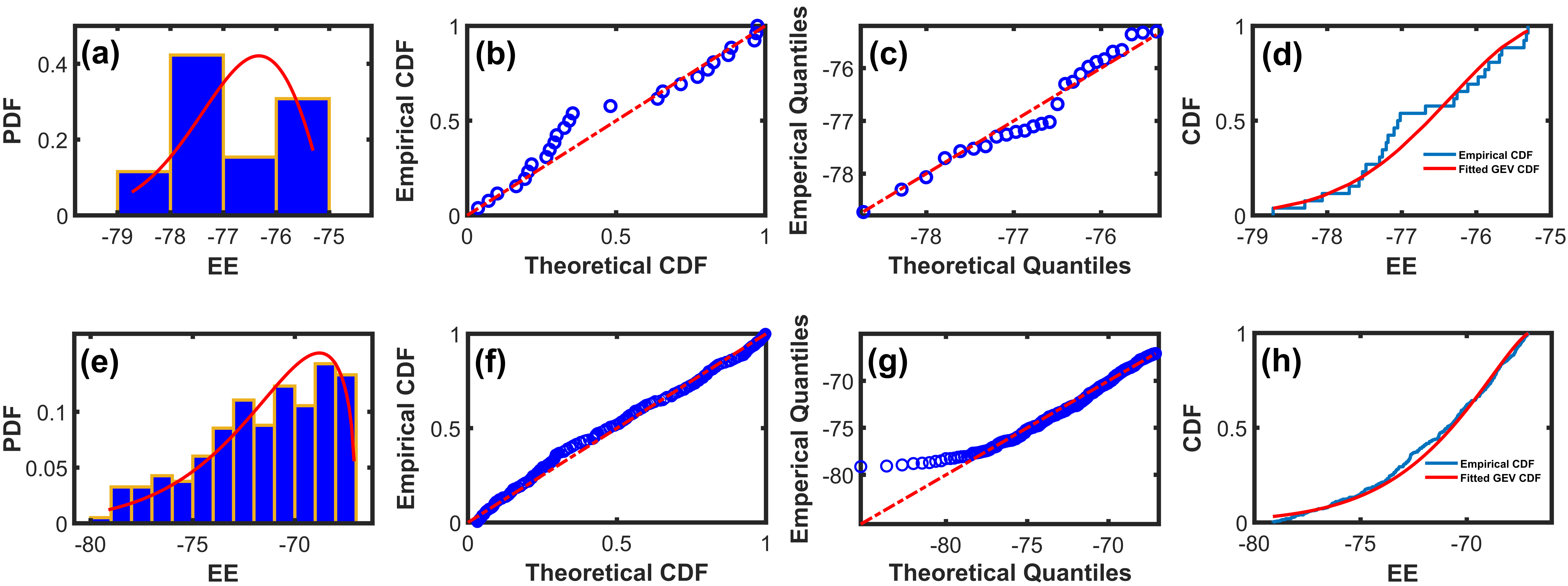}
	%\captionsetup{justification=justified} % Caption justified in two columns
	\caption{\textbf{Probability density function and test of goodness of fit:} The probability density function, P-P plot, Q-Q plot, and K-S statistic plot of all the EEs for $k\approx0.1$ are represented in (a)-(d), respectively. (e)-(h) Delineation of the probability density function, P-P plot, Q-Q plot, and K-S statistic plot of EEs for $k\approx0.06$ respectively.
	}
	\label{ee}
\end{figure*}

The statistics of EEs, i.e., the \textit{events}, which are being transcended by the threshold $H_{th}$ line for $k\approx0.1$ and $k\approx0.06$ are presented in Fig.~\ref{ee}. The probability density functions (PDF) of the EEs for $k\approx0.1$ and $k\approx0.06$ are demonstrated in Figs.~\ref{ee}(a) and \ref{ee}(e), respectively. We find that in both cases the PDFs fit well with the GEV distribution prescribed by the following mathematical form, 
\begin{eqnarray}
	G(x) = \dfrac{1}{\beta}\exp\Bigg(-\Big(1+\gamma\dfrac{x-\alpha}{\beta}\Big)^{-\frac{1}{\gamma}}\Bigg) \times \Big(1+\gamma\dfrac{x-\alpha}{\beta}\Big)^{-\frac{1}{\gamma}-1} \nonumber \\
	\label{dis1}
\end{eqnarray}
for $\beta\neq0$ and $1+\gamma\dfrac{x-\alpha}{\beta} > 0$. Here $\alpha>0$ is location parameter, $\beta >0$ is scale parameter, and $\gamma\neq0$ signifies the shape parameter. Depending on whether $\gamma$ is positive ( $\gamma>0$) or negative ($\gamma<0$) the distribution type varies as Fr\'echet (type II) and Weibull (type III) distributions respectively. There is a third case scenario where the shape parameter $\gamma=0$. Such distributions are called Gumbell (type I) and in this case, the distribution takes the following mathematical form,
\begin{eqnarray}
	G(x) = \dfrac{1}{\beta}\exp\Bigg(-\exp\Bigg(-\dfrac{x-\alpha}{\beta}\Bigg)-\dfrac{x-\alpha}{\beta}\Bigg).
	\label{dis2}
\end{eqnarray}
The performance of the K-S test for both sets of EEs corresponding to $k\approx0.1$ and $k\approx0.06$ fails to reject the \textit{null hypothesis} (the data sets follow the GEV distribution) in the $95\%$ confidence interval, which substantiates that both the data sets follow the GEV distribution. The details of the K-S test result are evinced in table \ref{table1}.
\begin{table} 	
	\begin{tabular}{|c|c|c|c|}
		\hline%\hline
		\multicolumn{4}{|c|}{K-S Statistics of GEV distributions of EEs } \\\hline\hline
		Data & Coupling Strength & Parameter & Estimate \\ 
		\hline\hline
		& & $p$-value & $0.3024$ \\ 
		\cline{3-4}
		& $k\approx$ 0.1 & & \\ 
		\cline{3-4}
		& & $K$-$S$ Statistics & $0.1842$ \\
		\cline{2-4}
		EE & & & \\
		\cline{2-4}
		& & $p$-value & $0.0671$ \\ 
		\cline{3-4}
		& $k\approx$ 0.06 & &\\ 
		\cline{3-4}
		& & $K$-$S$ Statistics & $0.0649$ \\ 
		\hline%\hline
	\end{tabular}
	\caption{ K-S test results of the sets of EEs concerning $k\approx0.1$ and $k\approx0.06$, respectively. }
	\label{table1}
\end{table}

 As the visual attestation of the K-S test, the empirical cumulative distribution function (CDF), i.e., the CDF, which indeed is calculated based on EEs data gathered from the numerical simulation and the CDF being calculated from the fitted GEV distribution curve of the numerically simulated EEs data sets, are presented by the blue and the red curves, respectively, in the respective figures Figs.~\ref{ee}(d) and \ref{ee}(h) for the respective sets of EEs corresponding to $k\approx0.1$ and $k\approx0.06$. The details of the GEV distribution curves of the EEs, i.e., the parameter values, confidence intervals ($95\%$) of the parameters, and the standard errors of the parameters are shown in table \ref{table2}. In both cases, the shape parameter ($\gamma$) values of the GEV distributions being negative, the distributions are intrinsically \textit{Weibull}. For the visual affirmation of how good these GEV distribution fittings are, the probability-probability plot (P-P plot) and quantile-quantile plot (Q-Q plot) are demonstrated in Figs.~\ref{ee}(b) and \ref{ee}(c) for the set of EEs corresponding to $k\approx0.1$  and in Figs.~\ref{ee}(f) and \ref{ee}(g) for the set of EEs regarding $k\approx0.06$.
\begin{table*} 	
	\begin{tabular}{|c|c|c|c|c|c|}
		\hline%\hline
		\multicolumn{6}{|c|}{Statistics of the GEV distribution} \\\hline\hline
		Data & Coupling Strength & Parameter & Estimate & Confidence Interval ($95\%$)& Standard Error \\ 
		\hline\hline
		& & $\gamma$ & $-0.53247$ & $[-1.0083, -0.056588]$ & $0.2428$ \\ 
		\cline{3-6}
		& $k\approx$ 0.1 & $\beta$ & $1.0432$ & $[0.69854, 1.5579]$ & $0.21923$ \\ 
		\cline{3-6}
		& & $\alpha$ & $-76.982$ & $[-77.4579, -76.506]$ & $0.24282$ \\
		\cline{2-6}
		EE & & & & &\\
		\cline{2-6}
		& & $\gamma$ & $-0.74396$ & $[-0.8221, -0.66582]$ & $0.039869$ \\ 
		\cline{3-6}
		& $k\approx$ 0.06 & $\beta$ & $3.574$ & $[3.2446, 3.9368]$ & $0.17658$ \\ 
		\cline{3-6}
		& & $\alpha$ & $-71.8571$ & $[-72.2343, -71.4799]$ & $0.19246$ \\ 
		\hline%\hline
	\end{tabular}
     \caption{Parameter values of the GEV distributions of the sets of EEs concerning $k\approx0.1$ and $k\approx0.06$.}
     \label{table2}
\end{table*}
 
\begin{figure*}[!ht]
	\centering
	\includegraphics[width=1.0\textwidth]{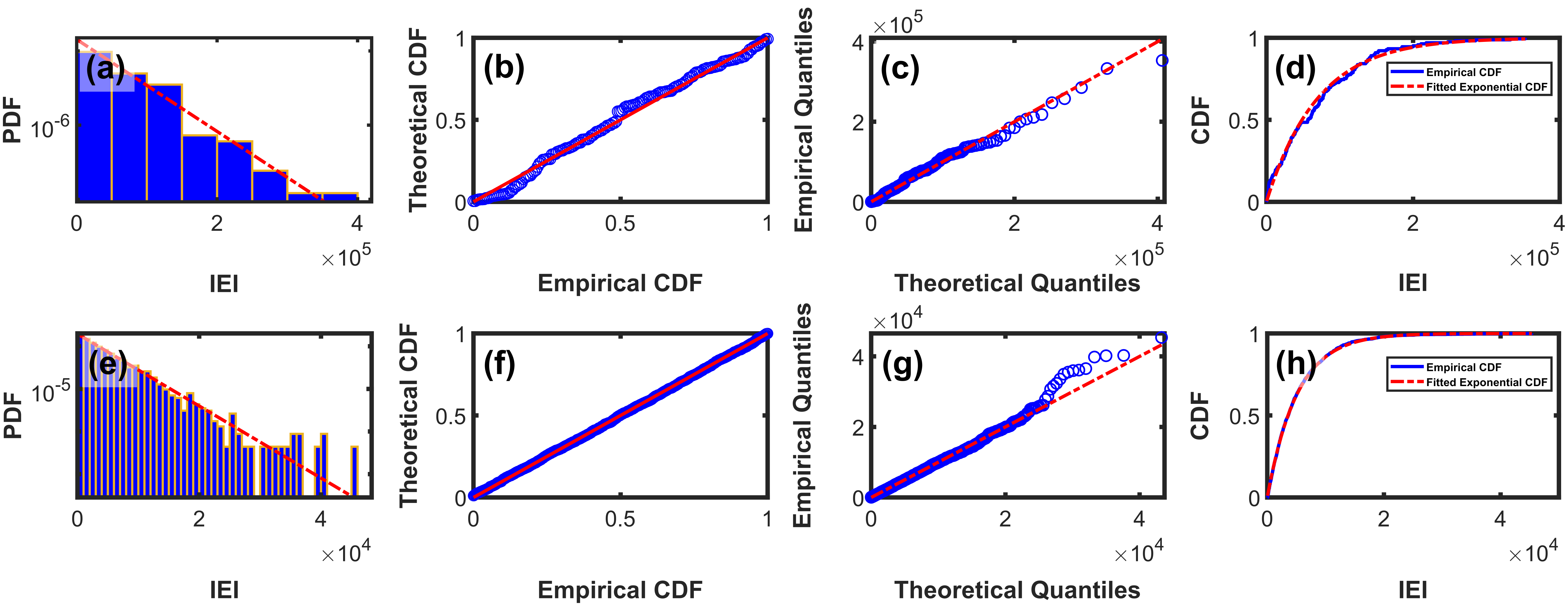}
	%\captionsetup{justification=justified} % Caption justified in two columns
	\caption{\textbf{Statistics of IEIs:} (a)-(d) The probability density function, P-P plot, Q-Q plot, and the K-S statistic of the IEIs for $k\approx0.1$ are displayed, respectively. The probability density function, P-P plot, Q-Q plot, and the K-S statistics of IEIs for $k\approx0.06$ are portrayed in (e)-(h), respectively.
	}
	\label{iei}
\end{figure*}

The time elapsed in forming two successive EEs is enunciated as an \textit{inter-event interval} (IEI). The statistics of IEIs for $k\approx0.1$ and $k\approx0.06$ are presented in Fig. \eqref{iei}. In the first panel, Figs.~\ref{iei}(a) and \ref{iei}(e) represent the PDF of sets of IEIs corresponding to $k\approx0.1$ and $k\approx0.06$, respectively. Both the PDFs fit well with \textit{exponential} distribution, narrated by the following mathematical form,
\begin{equation}\label{eq.5}
	\begin{array}{lcl} f(x;\mu)=
		\begin{cases} 
			\mu e^{-\mu x}; &  ~x \ge 0,  \\
			0; & ~x <0,
		\end{cases}
	\end{array}
\end{equation}

where $\mu >0$ describes the parameter. The details of the \textit{exponential} distribution curves is presented in table \ref{table3}.
\begin{table*} 	
	\begin{tabular}{|c|c|c|c|c|c|}
		\hline%\hline
		\multicolumn{6}{|c|}{Statistics of the exponential distribution} \\\hline\hline
		Data & Coupling Strength & Parameter & Estimate & Confidence Interval ($95\%$)& Standard Error \\ 
		\hline\hline
		& $k\approx$ 0.1 & $\mu$ & $70054.7$ & $[60483.7, 82104.7]$  & 5515.55 \\ 
		\cline{2-6}
		IEI & & & & &\\
		\cline{2-6}
		& $k\approx$ 0.06 & $\mu$ & $-71.8571$ & $[-72.2343, -71.4799]$ & 107.76 \\ 
		\hline%\hline
	\end{tabular}
	\caption{Parameter values of the exponential distributions of the sets of IEIs relating to $k\approx0.1$ and $k\approx0.06$.}
	\label{table3}
\end{table*}
K-S test analysis for the goodness of fit of the \textit{exponential} distribution for the sets of IEIs relating to $k\approx0.1$ and $k\approx0.06$ fail to reject the \textit{null hypothesis}, affirming the data sets follow the \textit{exponential} distribution. The details of the K-S test analysis are presented in table \ref{table4}. In the second panel, Figs.~\ref{iei}(b) and \ref{iei}(f) show the P-P plots, and in the third panel, Figs.~\ref{iei}(c) and \ref{iei}(g) display Q-Q plots of the sets of IEIs for $k\approx0.1$ and $k\approx0.06$, respectively. Figures ~\ref{iei}(d) and \ref{iei}(h) are concerning K-S tests for the respective sets of IEIs relating to $k\approx0.1$ and $k\approx0.06$; the \textit{blue} curve countenance of the \textit{empirical} CDF, which is basically calculated based on the data sets of IEIs accumulated from the numerical simulation; and the \textit{red} curve signify the \textit{theoretical} CDF being calculated from the fitted \textit{exponential} distribution curve regarding the sets of IEIs corresponding to $k\approx0.1$ and $k\approx0.06$, respectively.
\begin{table} 	
	\begin{tabular}{|c|c|c|c|}
		\hline%\hline
		\multicolumn{4}{|c|}{K-S Statistics of exponential distributions of IEIs } \\\hline\hline
		Data & Coupling Strength & Parameter & Estimate \\ 
		\hline\hline
		& & $p$-value & $0.4414$ \\ 
		\cline{3-4}
		& $k\approx$ 0.1 & & \\ 
		\cline{3-4}
		& & $K$-$S$ Statistics & $0.0664$ \\
		\cline{2-4}
		IEI & & & \\
		\cline{2-4}
		& & $p$-value & $0.7110$ \\ 
		\cline{3-4}
		& $k\approx$ 0.06 & &\\ 
		\cline{3-4}
		& & $K$-$S$ Statistics & $0.0146$ \\ 
		\hline%\hline
	\end{tabular}
	\caption{ K-S test results of the sets of IEIs concerning $k\approx0.1$ and $k\approx0.06$, respectively. }
	\label{table4}
\end{table}

\section{Mechanism behind the genesis of extreme events}
\label{sec:6}
\subsection{Occasional in-phase synchronization} 
\label{subsec:6a}
\begin{figure*}[!ht]
	\centering
	\includegraphics[width=1.0\linewidth]{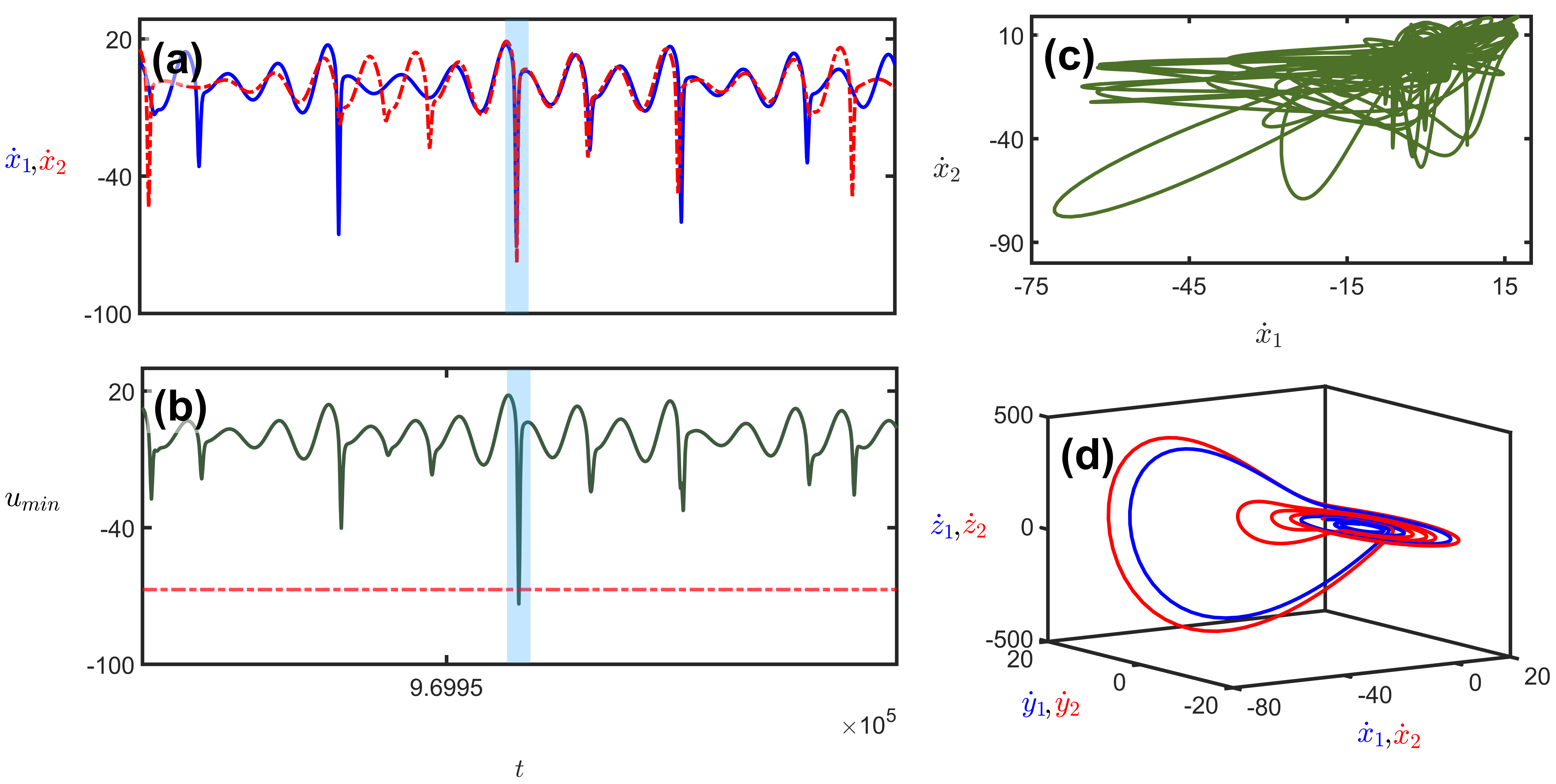}
	%\captionsetup{justification=justified} % Caption justified in two columns
	\caption{\textbf{Occasional in-phase synchronization:} (a) Temporal evolution of the individual oscillators of the system \eqref{mastereqn} for $\dot{x}_1$ (solid blue) and $\dot{x}_2$ (dotted red) corresponding to the coupling strength $k\approx0.06$. The shaded \textit{cyan} portion depicts the \textit{in-phase} synchronization region. (b) The temporal dynamics of the observable $u$ concerning $k\approx0.06$. The \textit{cyan-colored} shaded spike confirms the extreme event in the $u$ variable by exceeding the \textit{red-dashed} threshold, $H_{th}=\mu-5\sigma$ ($\mu$ is the mean and $\sigma$ is the standard deviation of the set of local minima of the observable $u$), line. (c) The phase portrait of the system \eqref{mastereqn} upon $(\dot{x}_{1},\dot{x}_{2})$ plane concerning the time series (a). The largely deviated trajectory is along the \textit{in-phase} synchronization manifold. (d) The phase portrait of the system \eqref{mastereqn} regarding the time series (a) on the $3-dimensional$ $(\dot{x},\dot{y},\dot{z})$ plane.
	}
	\label{phase}
\end{figure*}

So far, we have reported the advent of \textit{extreme events} in the system \eqref{mastereqn} and described a detailed statistical analysis of EEs in the $x-directional$ average velocity variable $u$. This section claims its exigency by reckoning how the EEs are being originated in the very system, emphasizing the detailed dynamical procedure. Interestingly, we find the mechanism behind the emergence of EEs is the \textit{occasional in-phase} synchronization. The \textit{occasional in-phase} synchronization is a phenomenon that occurs in coupled systems wherein two oscillators, evolving asynchronously each other, occasionally visit the \textit{in-phase} synchronization manifold (SM). During the occasional visit to \textit{in-phase} SM, the individual oscillators coincide phase-on-phase with each other, during which extreme events occur in the coupled system. A prominent exposition of this mechanism correlating our study is presented in Fig.~\ref{phase}. The temporal evolution of the individual oscillators for $k\approx0.06$ of the coupled system \eqref{mastereqn}, $\dot{x}_1$ (solid blue) and $\dot{x}_2$ (dotted red) are plotted one over the other in Fig.~\ref{phase}(a).    
 %\begin{widetext}
%\end{widetext}
\begin{figure*}[!ht]
	\centering
	\includegraphics[width=1.0\linewidth]{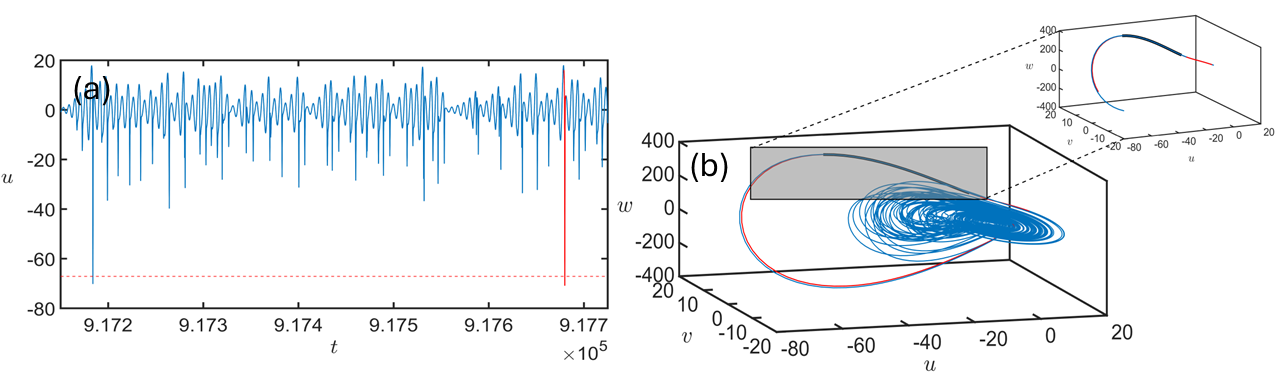}
	%\captionsetup{justification=justified} % Caption justified in two columns
	\caption{\textbf{Depiction of channel-like structure:} (a) The time series of the observable $u$ for $k\approx0.06$. Two large amplitude spikes are shown to exceed the \textit{extreme event} qualifying threshold $H_{th}$ line (\textit{red-dashed}). (b) The phase portrait of \eqref{mastereqn} corresponds to the time series (a) upon the $(u,v,w)$ plane.
	}
	\label{channel}
\end{figure*}
%\end{widetext}

Figure ~\ref{phase}(b) is the presentation of the temporal dynamics of the observable $u$. The \textit{cyan} shaded region in Fig.~\ref{phase}(a) illustrates the \textit{in-phase} synchronization of the temporal dynamics of the variables $\dot{x_{1}}$ and $\dot{x_{2}}$. Due to this \textit{in-phase} synchronization, the large excursion of the observable $u$ is produced, and that is apotheosized by the \textit{cyan} shaded region of the temporal evolution of $u$ presented in Fig.~\ref{phase}(b). Figure~\ref{phase}(c) is the depiction of the phase portrait upon the $(\dot{x_{1}},\dot{x_{2}})$ plane corresponding to the time series presented in Fig.~\ref{phase}(a). The large aberration of the trajectory in the phase portrait is along the \textit{in-phase} direction. Figure~\ref{phase}(d) is the presentation of the phase portrait of the individual oscillator concerning the time series presented in Fig.~\ref{phase}(a) in the $3-dimensional$ plane ($\dot{x},\dot{y},\dot{z}$), where the \textit{blue} trajectory corresponds to the first oscillator and the \textit{red} trajectory corresponds to the dynamics of the second oscillator.

\textit{Channel-like structure}: Meticulous observation of the dynamics concerning how the EEs are developed in the variable $u$ of the system \eqref{mastereqn} unveils another new era. Interestingly, corresponding to the high amplitude spikes categorized as EEs in the temporal evolution of $u$, the largely deviated trajectories upon the $3-dimensional$ phase space $(u,v,w)$ initiate by passing through a small \textit{channel-like} structure. This might be a considerable potential mechanism behind the emergence of EEs in the variable $u$. As an illustration in Fig.~\ref{channel}(a), the temporal dynamics of $u$ corresponding to the coupling strength $k\approx0.06$ is displayed. Two \textit{extreme} trajectories (\textit{blue} and \textit{red}) are depicted. The phase portrait of the system \eqref{mastereqn} upon $(u,v,w)$ plane concerning the time series presented in Fig.~\ref{channel}(a) is delineated in Fig.~\ref{channel}(b). It is quite discernible that the two large excursions ratifying \textit{extreme events} initiate proceeding through a small \textit{channel-like} structure shown by the shaded region in the inset figure of Fig.~\ref{channel}(b).

\subsection{On-off intermittency}
\label{subsec:6b}
\textit{Synchronization error dynamics}: The presence of heterogeneity in the frequency parameter (i.e., $\omega_{1}\ne\omega_{2}$) persuades our attention to investigate the dynamical behavior of the synchronization error (SE) dynamics of the system \eqref{mastereqn}. The SE is defined by the following mathematical expression.
\begin{figure}[!ht]
	\centering
	\includegraphics[width=1.0\linewidth]{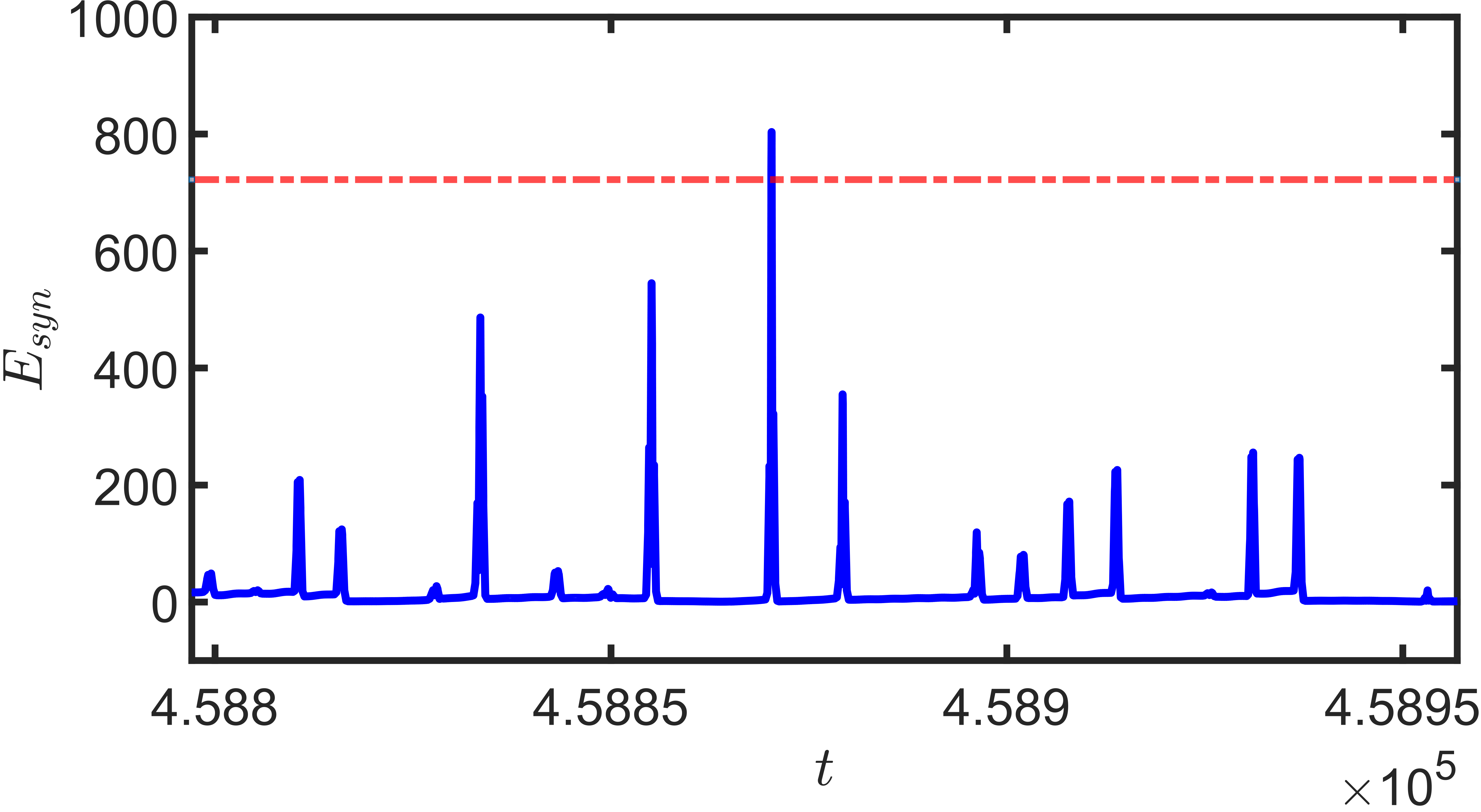}
	\caption{\textbf{Temporal dynamics of the synchronization error:} The time series of the synchronization error ($E_{syn}$) for the coupling strength $k\approx0.035$ is shown in the figure. On-off type intermittency is observed. The \textit{red-dashed} line represents the threshold $H_{th}=\mu+6\sigma$, where $\mu$ is the mean and $\sigma$ is the standard deviation of the set of $E_{syn}$. 
	}
	\label{syncerr}
\end{figure}

\begin{figure}[!ht]
	\centering
	\includegraphics[width=0.9\linewidth]{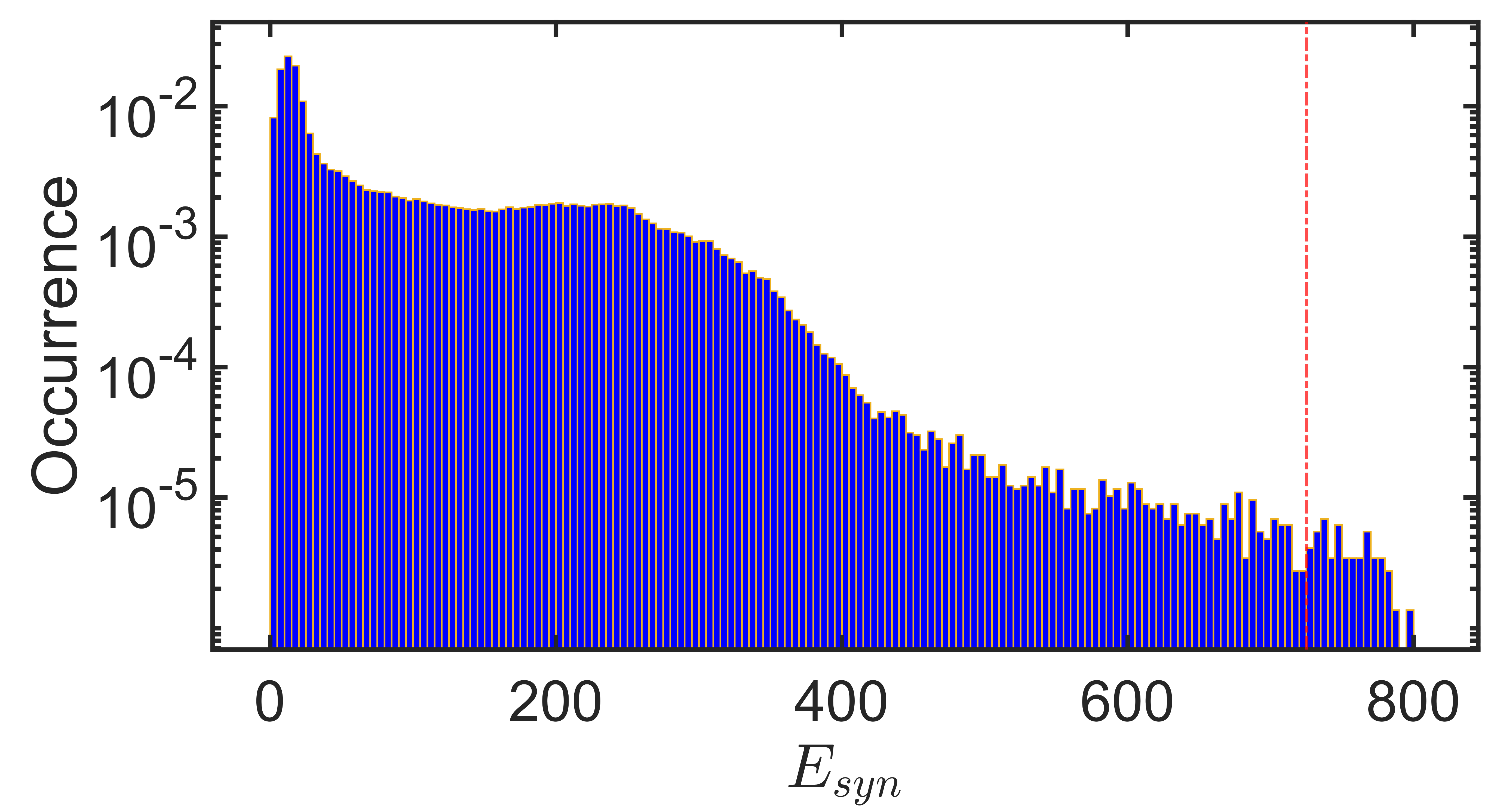}
	\caption{\textbf{Histogram of the synchronization error ($E_{syn}$):} The histogram for the st of $E_{syn}$ concerning $k\approx0.035$ is presented here. The \textit{red} vertical dashed line stands for the threshold $H_{th}=\mu+6\sigma$, where $\mu$ is the mean and $\sigma$ is the standard deviation of the set of $E_{syn}$. The portion of the histogram that is beyond the vertical threshold line towards the right indeed corroborates the extreme events. This region is generally contemplated as the tail of the histogram. 
	}
	\label{errsync_hist}
\end{figure}

\begin{eqnarray}
	E_{syn} = \langle\sqrt{(\dot{x}_1-\dot{x}_2)^2+(\dot{y}_1-\dot{y}_2)^2+(\dot{z}_1-\dot{z}_2)^2}\rangle_t.
\end{eqnarray}
As an archetypal example for illustration, a time series of SE concerning $k\approx0.035$ is presented in Fig.~\ref{syncerr}. Interestingly, we observe a few large amplitude spikes in the time series following \textit{on-off} intermittency-like behavior. One such spike is shown in the temporal evolution to surpass the \textit{red-dashed} threshold line $H_{th}=\mu+6\sigma$, where $\mu$ and $\sigma$ are the mean and standard deviation of the set of $E_{syn}$ for $k\approx0.035$. The large amplitude spikes (SE), which surpass the $H_{th}$ line, are regarded as \textit{extreme events} in synchronization error dynamics. The histogram of $E_{syn}$ for the coupling strength $k\approx0.035$ is plotted in Fig.~\ref{errsync_hist}. The \textit{red} vertical dashed line represents the threshold $H_{th}$. The portion of the histogram that is at the right of the vertical threshold line basically appears for the \textit{extreme events}. This region is also contemplated as the tail of the histogram. 
\begin{figure}[!ht]
	\centering
	\includegraphics[width=1.0\linewidth]{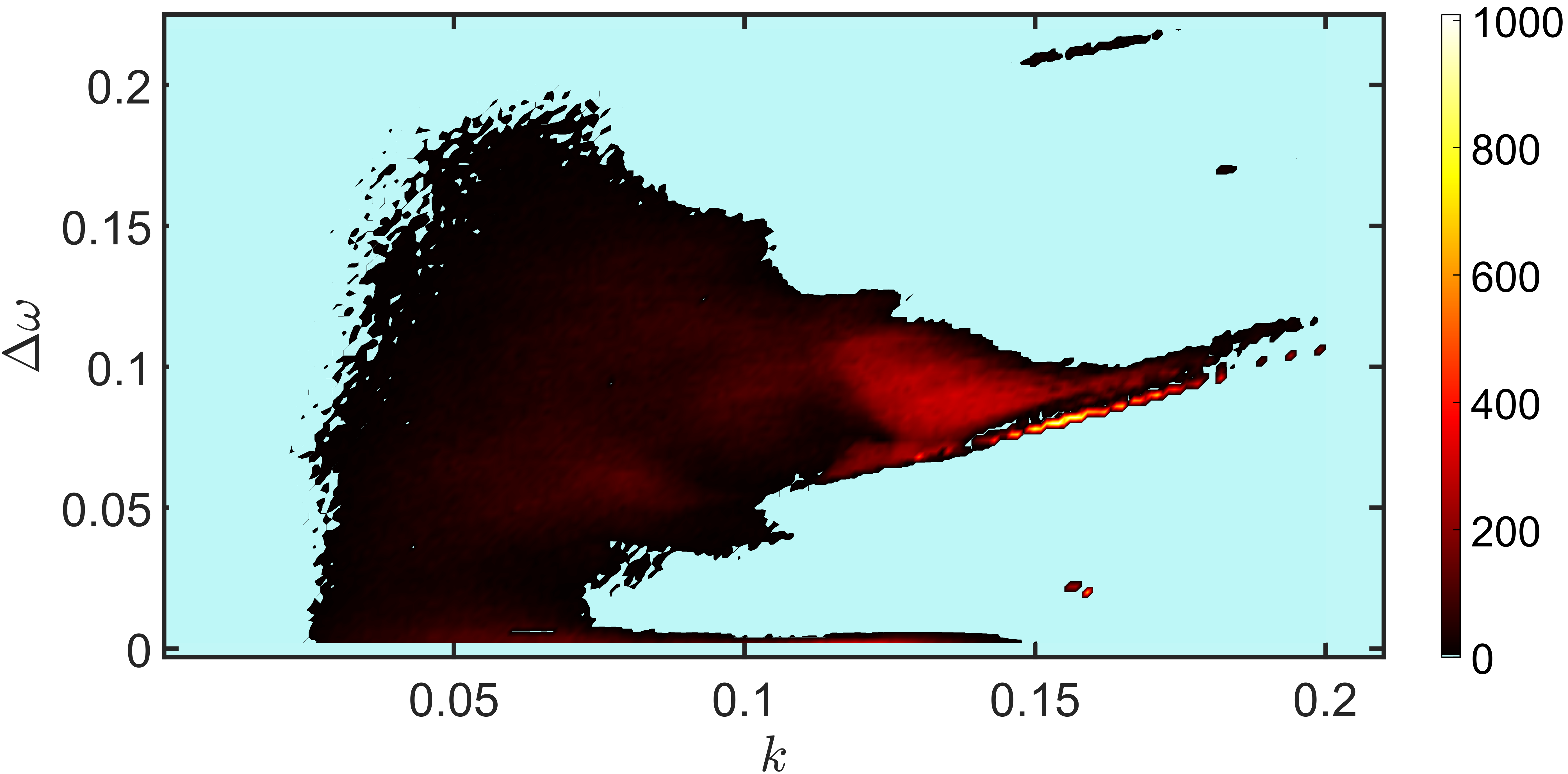}
	\caption{\textbf{Parameter space explicating the region of the emergence of extreme events in the synchronization error ($E_{syn}$) dynamics upon $(k,\Delta\omega)$ plane:} The \textit{cyan} region stands for the non-extreme events region, and the colored portion represents the extreme events region.}
	\label{syncerr_paraspace}
\end{figure}
To corroborate the region of the emergence of \textit{extreme events} in synchronization error dynamics, a parameter space for the system \eqref{mastereqn} on the $(k,\Delta\omega)$ plane is plotted in Fig.~\ref{syncerr_paraspace}. The \textit{cyan} region represents the non-extreme events region, and the \textit{other-colored} region represents the extreme events region.

\textit{Transverse direction to the synchronization manifold}: For a system of coupled oscillators, the synchronization state is the utmost expedient, but due to the presence of noise or system parameter heterogeneity, this synchronized state might be impeded. A stereotypical scenario is espied for the trajectory of the coupled system, that it experiences an excursion from the \textit{synchronization manifold} towards the \textit{transverse} direction for a short while and follows a return back due to the nonlinear folding of the flow. This evanescent and intermittent excursion of the trajectories is contemplated as the bubbling of the attractor. The two-coupled system \eqref{mastereqn} lives in a six-dimensional ($6D$) phase space spanned by $(x_{i},y_{i},z_{i}), i=1,2,...,6$. 
\begin{figure}[!ht]
	\centering
	\includegraphics[width=1.0\linewidth]{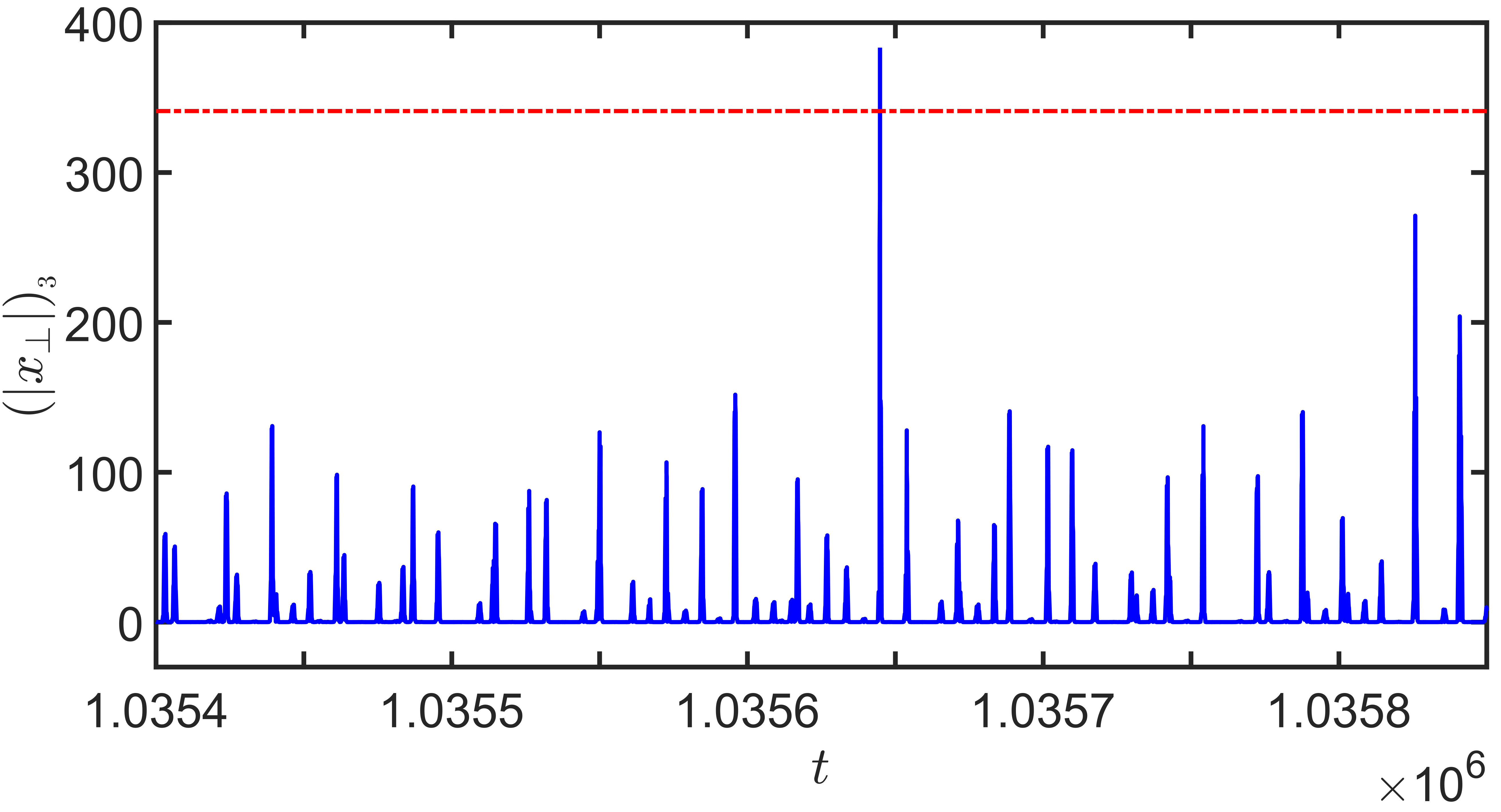}
	\caption{\textbf{Temporal dynamics of the variable towards the transverse direction of the synchronization manifold:} The temporal evolution of the $(x_{\perp})_{3}$ variable is presented here for the coupling strength $k\approx0.06$. On-off type intermittency is clearly conspicuous in the time series. The \textit{red} horizontal dashed line represents the \textit{extreme events} qualifying threshold line, $H_{th}=\mu+6\sigma$, where $\mu$ and $\sigma$ are the mean and standard deviation of the set of $(x_{\perp})_{3}$.
	}
	\label{transversex3}
\end{figure}

In the case of synchronization, the two-coupled system resides in a $3D$ subspace (synchronization manifold). In this case new $3D$ state vectors are introduced: $(x_{\parallel})_{1}=\frac{\dot{x_{1}}+\dot{x_{2}}}{2}$, $(x_{\parallel})_{2}=\frac{\dot{y_{1}}+\dot{y_{2}}}{2}$, and $(x_{\parallel})_{3}=\frac{\dot{z_{1}}+\dot{z_{2}}}{2}$, pronouncing the behavior on the synchronization manifold, and $(x_{\perp})_{1}=\frac{\dot{x_{1}}-\dot{x_{2}}}{2}$, $(x_{\perp})_{2}=\frac{\dot{y_{1}}-\dot{y_{2}}}{2}$, and $(x_{\perp})_{3}=\frac{\dot{z_{1}}-\dot{z_{2}}}{2}$ elucidating the behavior along the transverse direction to the synchronization manifold. 

\begin{figure}[!ht]
	\centering
	\includegraphics[width=1.0\linewidth]{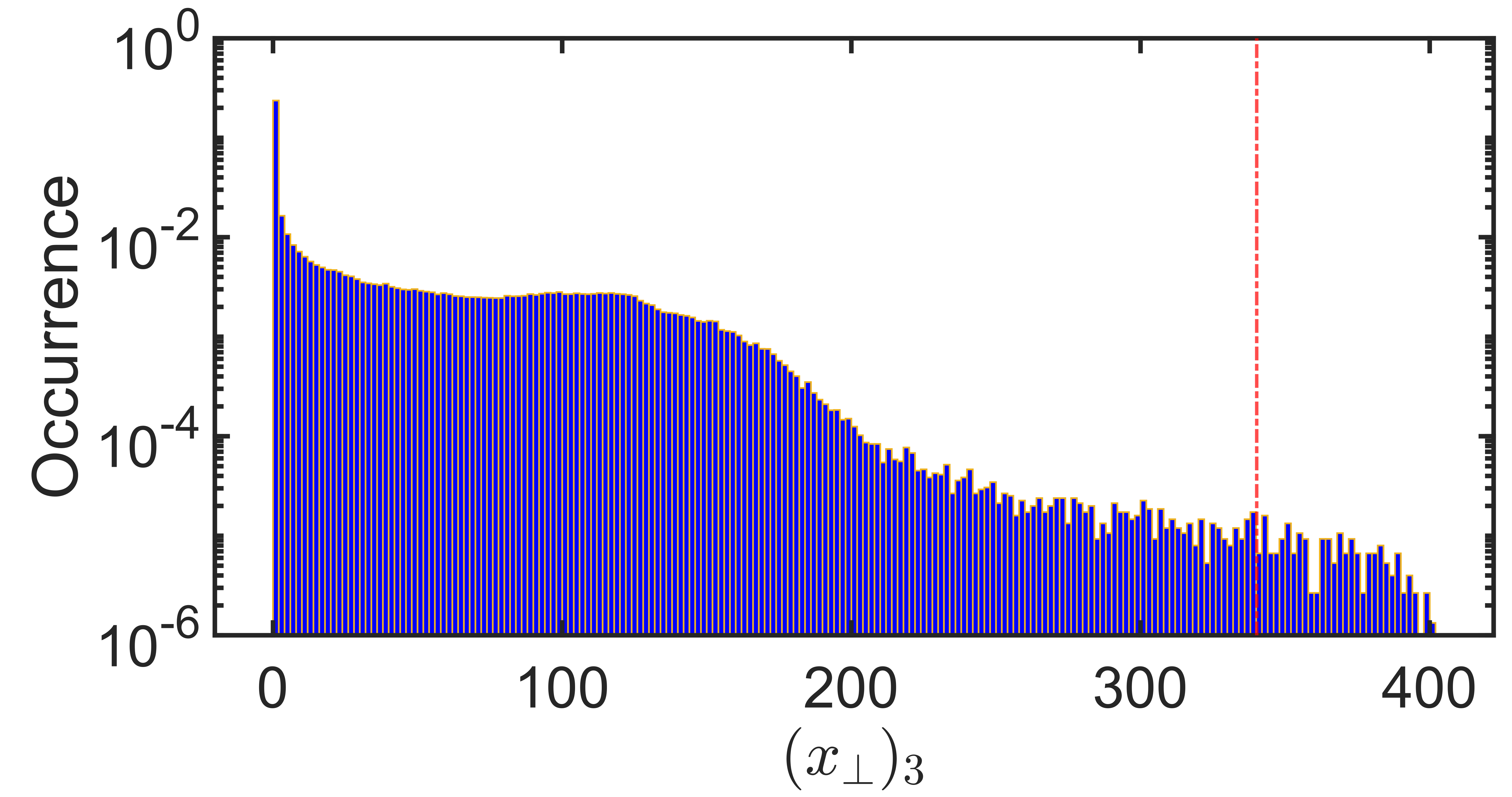}
	\caption{\textbf{Histogram of $(x_{\perp})_{3}$:} The histogram of the set of data comprising of $(x_{\perp})_{3}$ for the coupling strength $k\approx0.06$ is presented in this figure. The \textit{red} vertical dashed line represents the threshold $H_{th}=\mu+6\sigma$, where $\mu$ is the mean and $\sigma$ is the standard deviation of the set of $(x_{\perp})_{3}$. The right side portion of the histogram to the vertical threshold line is the tail of the histogram.}
	\label{xper3_hist}
\end{figure}

Figure~\ref{transversex3} represents the temporal dynamics of $(x_{\perp})_{3}$ state vector for $k\approx0.06$. In the temporal evolution, a few high-amplitude chaotic bursts are observed, among which one high-amplitude spike is discriminated to exceed the \textit{red-dashed} threshold line, $H_{th}=\mu+6\sigma$, where $\mu$ and $\sigma$ are the mean and the standard deviation of the set of local maxima of $(x_{\perp})_{3}$, respectively. The histogram of the set of $(x_{\perp})_{3}$ corresponding to the coupling strength $k\approx0.06$ is presented in Fig.~\ref{xper3_hist}. The \textit{red} vertical dashed line depicts the threshold $H_{th}$. The region of the histogram of the right side of the threshold line corroborates the \textit{extreme events}. This portion is also contemplated as the tail of the histogram.

 \begin{figure}[!ht]
	\centering
	\includegraphics[width=1.0\linewidth]{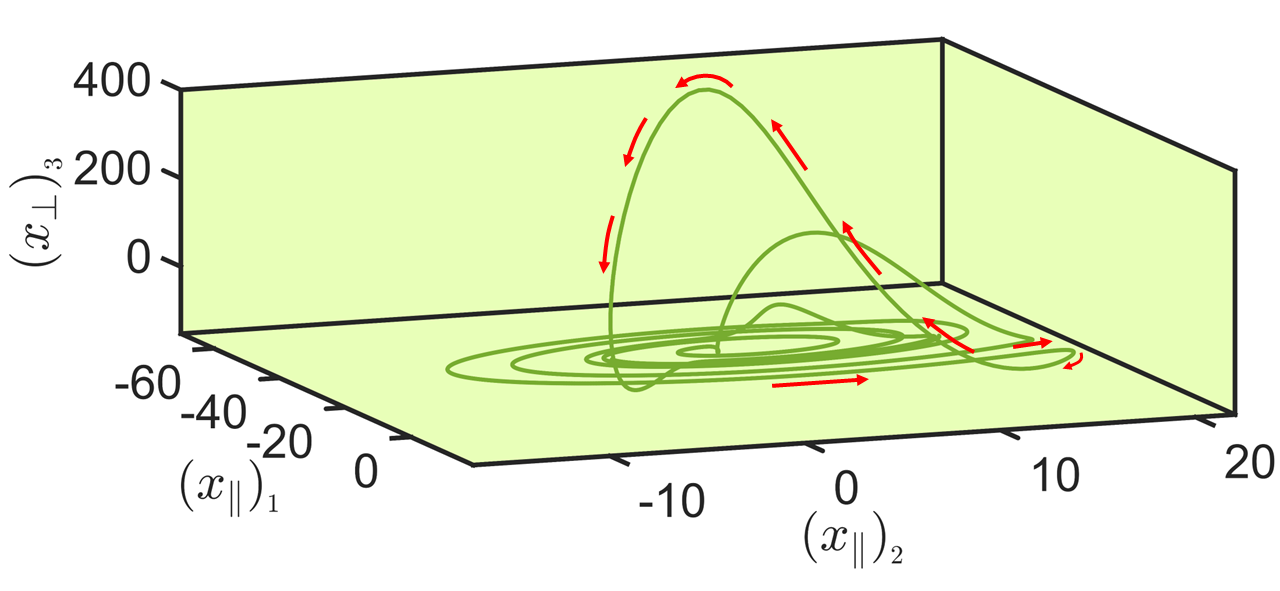}
	\caption{\textbf{Bubble transition:} Bubbling of the attractor corresponding to the time series presented at Fig.~\ref{transversex3} for the coupling strength $k\approx0.06$ is shown here. The trajectories mostly circulate, residing on the invariant manifold; occasionally they traverse in the transverse direction to the synchronization manifold as large excursions. The \textit{red-arrow} shaded large bubble concerns the large amplitude \textit{extreme event} in the time series.}
	\label{bubbling}
\end{figure}
Here the large amplitude spikes that cross the \textit{red-dashed} threshold $H_{th}$ line are considered as \textit{extreme events}. Corresponding to the \textit{extreme event} shown as the large amplitude spike crossing the \textit{red-dashed} threshold line presented in Fig.~\ref{transversex3}, the typical \textit{bubble} as the projection of $6D$ phase space upon $3D$ phase space containing the components of the \textit{synchronization manifold} and of the \textit{transverse manifold} is portrayed in Fig.~\ref{bubbling}.
\begin{figure}[!ht]
	\centering
	\includegraphics[width=1.0\linewidth]{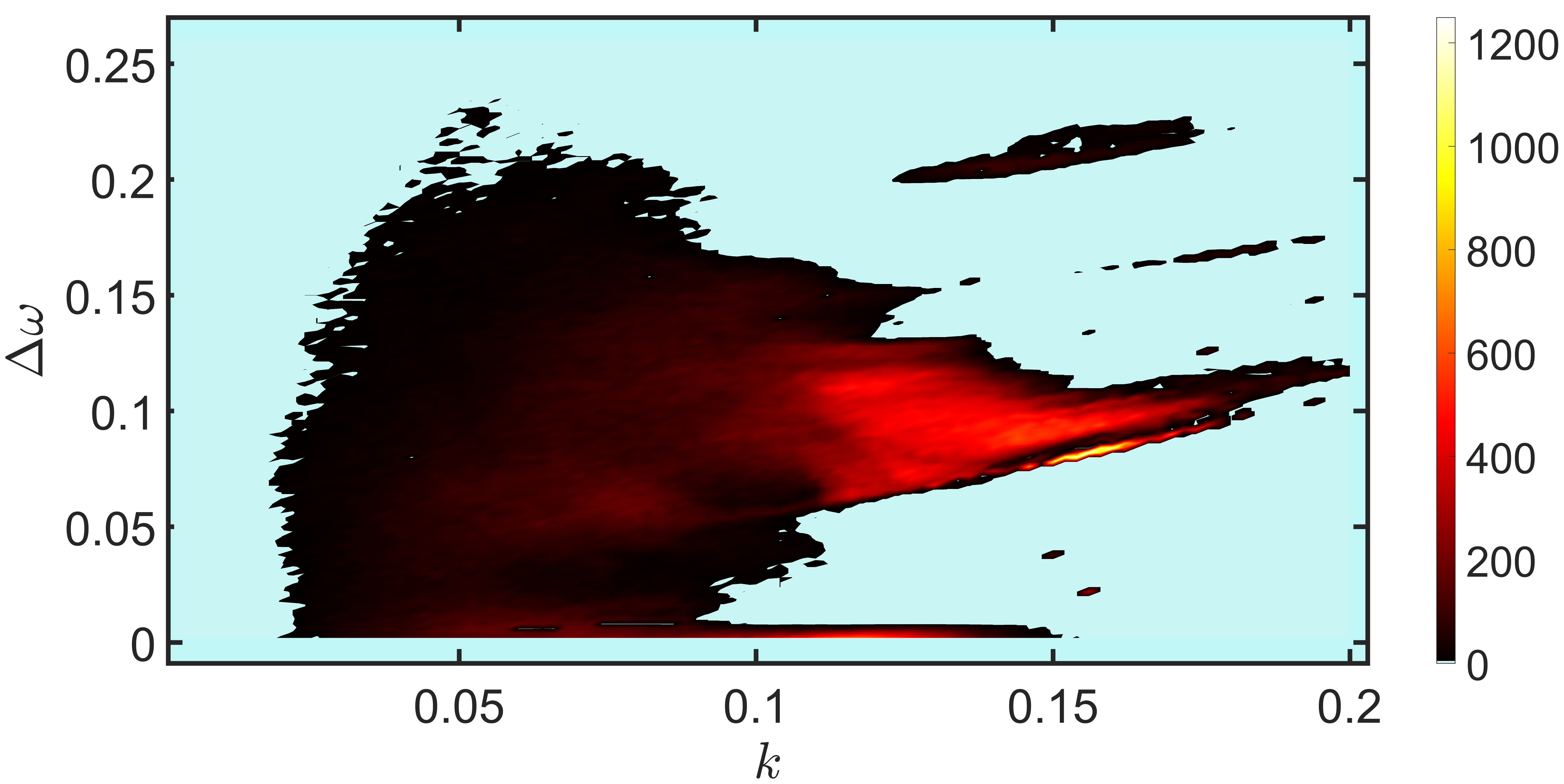}
	\caption{\textbf{Parameter space delineating the region of the emergence of extreme events of the variable $(x_{\perp})_{3}$ upon $(k,\Delta\omega)$ plane:} The \textit{cyan} region depicts the non-extreme events region, and the colored region represents the \textit{extreme events} region.}
	\label{trans_xperp3_paraspace}
\end{figure}
To substantiate the emerging region of extreme events in the state vector $(x_{\perp})_{3}$, a parameter space upon the $(k,\Delta\omega)$ plane is displayed in Fig.~\ref{trans_xperp3_paraspace}. The \textit{cyan} region elucidates the non-extreme event region, and the \textit{colored} region confirms the extreme event emergence region.

\section{Conclusion}
\label{sec:8}
In this entrancing study, we have observed \textit{extreme events} and have unexplored the significant dynamical mechanisms leading to the emergence of EEs in a system of \textit{diffusively} and \textit{bidirectionally} two coupled R\"ossler oscillators. To classify the EEs, we have considered the threshold-based approach method. Interestingly, we have noticed the appearance of EEs in different observables: the average velocity variable ($u$) in the $x$ direction, the synchronization error dynamics, and one transverse directional variable ($(x_{\perp})_{3}$). \textit{Extreme events} emerge in the average velocity variable due to the occasional \textit{in-phase} synchronization of the respective velocity variables. In this case, the large excursions of the \textit{extreme} trajectories have noticed to begin by proceeding through a small \textit{channel-like} structure. The underlying mechanism behind the origination of EEs in synchronization error and transverse directional variable dynamics is \textit{on-off intermittency}. We have also noticed the bubbling of the chaotic attractor regarding the \textit{on-off intermittency} in the $(x_{\perp})_{3}$ variable. We have performed the statistical analysis of the sets of EEs and the inter-event intervals (IEI) for the average velocity variable. The sets of EEs follow the GEV distribution and the sets of IEIs fits well with exponential distribution.
\par One intriguing research perspective might be the finding, whether there is any interrelation between the \textit{on-off intermittency} and the Dragon King (DK) probability distribution of the events. The investigation of the emergence of EEs in different network configurations for the different number of R\"ossler oscillators might be a core boulevard of scientific research. One might be interested in finding whether these kinds of similar results are discernible in a system of \textit{diffusively} and \textit{bidirectionally} two coupled different types of three-dimensional chaotic oscillators. In the conclusion, we look for that our findings will embolden future research regarding R\"ossler oscillator, and our knowledge will contribute to finding a new era concerning the emergence of EEs in this oscillator.
  
\acknowledgments

S.S thanks the Science and Engineering Research Board (SERB), Department of Science and Technology (DST), Government of India for providing financial support in the form of National Post-Doctoral Fellowship (File No.~PDF/2022/001760). D.G. and T.K.P. are supported by Science and Engineering Research Board (SERB), Government of India (Project No. CRG/2021/005894). T.K.P. and S.S are humbly indebted to Prof. Syamal K. Dana for his sincere support and benevolent discussions.

\bibliography{ref.bib}
\end{document}